\pdfoutput=1
\documentclass[sigconf]{acmart}
\usepackage[utf8]{inputenc}
\usepackage{caption}
\usepackage{algorithm}
\usepackage{algpseudocode}
\usepackage{enumitem}
\usepackage{textcomp}
\usepackage{multirow}
\usepackage{balance}
\usepackage[inkscapelatex=false]{svg}

\usepackage{soul}
\usepackage{listings}
\usepackage[T1]{fontenc}
\newlength\mylen

\copyrightyear{2022} 
\acmYear{2022} 
\setcopyright{acmcopyright}
\acmConference[CCS '22]{Proceedings of the 2022 ACM SIGSAC Conference on Computer and Communications Security}{November 7--11, 2022}{Los Angeles, CA, USA}
\acmBooktitle{Proceedings of the 2022 ACM SIGSAC Conference on Computer and Communications Security (CCS '22), November 7--11, 2022, Los Angeles, CA, USA}
\acmPrice{15.00}
\acmISBN{978-1-4503-9450-5/22/11}
\acmDOI{10.1145/3548606.3560682}

\begin{document}

\title{Frequency Throttling Side-Channel Attack}

\author{Chen Liu}
\affiliation{
   \institution{Intel Corporation}
   \city{Hillsboro}
   \state{OR}
   \country{USA}}
\email{chen1.liu@intel.com}

\author{Abhishek Chakraborty}
\affiliation{
   \institution{Intel Corporation}
   \city{Hillsboro}
   \state{OR}
   \country{USA}}
\email{abhishek1.chakraborty@intel.com}

\author{Nikhil Chawla}
\affiliation{
   \institution{Intel Corporation}
   \city{Hillsboro}
   \state{OR}
   \country{USA}}
\email{nikhil.chawla@intel.com}

\author{Neer Roggel}
\affiliation{
   \institution{Intel Corporation}
   \city{Rio Rancho}
   \state{NM}
   \country{USA}}
\email{neer.roggel@intel.com}

\begin{abstract}
Modern processors dynamically control their operating frequency to optimize resource utilization, maximize energy savings, and conform to system-defined constraints.
If, during the execution of a software workload, the running average of any electrical or thermal parameter exceeds its corresponding predefined threshold value, the power management architecture will {\em reactively} adjust CPU frequency to ensure safe operating conditions.
In this paper, we demonstrate how such power management-based frequency throttling activity forms a source of  timing side-channel information leakage, which can be exploited by an attacker to infer secret data even from a constant-cycle victim workload.
The proposed frequency throttling side-channel attack can be launched by both kernel-space and user-space attackers, thus compromising security guarantees provided by isolation boundaries. We validate our attack methodology across different systems and threat models by performing experiments on a constant-cycle implementation of AES algorithm based on AES-NI instructions. The results of our experimental evaluations demonstrate that the attacker can successfully recover all bytes of an AES key by measuring encryption execution times.
Finally, we discuss different options to mitigate the threat posed by frequency throttling side-channel attacks, as well as their advantages and disadvantages.
\end{abstract}



\maketitle

\section{Introduction}\label{sec:intro}

\noindent Power management architectures of modern processor designs play a central role in optimizing for and balancing between high performance and low power consumption requirements, a product of decades of academic and industry innovation ~\cite{DBLP:journals/micro/RotemNAWR12, gough2015cpu, Intel-SDM, bircher2008analysis, energyefficiencybook, EnergyEfficientComputing}. For example, a widely-used power management architectural mechanism known as Dynamic Voltage and Frequency Scaling (DVFS) is available on Intel\textsuperscript{\textregistered}, AMD and ARM CPUs~\cite{kogler2022minefield,kim2008system}. DVFS dynamically adjusts CPU frequency and voltage in order to reduce system power consumption, yielding higher performance per Watt, or to quickly alter CPU frequency during workload execution, in order to ensure that different electrical and thermal parameters of the system remain below predefined safe limits~\cite{Intel-SDM, AMD-PB2, ARM-PM}. Similar such throttling has been recently identified as enabling covert channels~\cite{IChannels}, given its reliance on shared infrastructure across security domains. In this work, we investigate if such workload-dependent CPU frequency adjustments yield exploitable side-channels.

Modern systems also provide multiple software-accessible telemetries which allow users to characterize bottlenecks at scale ~\cite{AsmDB}, monitor resource utilization, power and performance ~\cite{MonitoringEnergyHotspots, Colmant2017WattsKitSP, Fieni2020SmartWattsSS}, and gain insights into system reliability ~\cite{MercurialCores}.
Recently, researchers have demonstrated how a processor's energy telemetry reporting framework can be used maliciously to perform power side-channel analysis attacks~\cite{Lipp2020Platypus,liu2021methodology}.
These attacks allow a user-space attacker (having Ring 3 privilege) to infer secret information from a targeted victim workload running inside a Trusted Execution Environment (TEE).
In order to thwart such side-channel attacks, CPU vendors (both Intel and AMD) have provided security patches~\cite{Intel-RAPL-Patch, AMD-RAPL-Patch} which remove Ring 3 software access to the energy telemetry data via the Linux kernel module. In addition, Intel has also provided a filtering-based mitigation patch~\cite{Intel-RAPL-Fuzz} to safeguard the reported energy telemetry readings even from a kernel-space attacker (having Ring 0 privilege).

In this paper, we study the potential threat posed by a new type of side-channel information leakage source termed the {\em frequency throttling side-channel}. Such a side-channel arises due to the dynamic adjustment of CPU frequency when workload execution causes one or more electrical or thermal system parameters to exceed predefined limits.
Typically, in such a scenario, the power management architecture throttles CPU frequency to a lower value to ensure safe operating conditions of the system. 
Then, depending on the running average of the parameter(s) in question, CPU frequency is again boosted to a higher value until any limit threshold is violated.
Therefore, increases and decreases in CPU frequency (hereinafter, referred to as {\em average throttling frequency}) during workload execution are dependent on the instantaneous electrical and thermal parameters being capped by system-defined limits.
The average throttled frequency in turn affects the overall execution time of the workload, even if its implementation follows constant-cycle coding principles~\cite{Intel-coding-guideline}, or it is being executed inside a TEE.
The objective of a side-channel attacker is to deduce the targeted secret from a victim workload by monitoring fluctuations in its execution time for processing different inputs. Unlike software-accessible telemetries, frequency throttling side-channel leakage cannot be thwarted by enforcing access restriction or filtering-based mitigation patches. This is because any presumed malicious Ring 3 software can precisely monitor the execution time of a targeted process by reading timestamp counter values (e.g., using the \emph{RDTSC} instruction).

In this work, we demonstrate the applicability of the throttling side-channel attack to retrieve secret information from cryptographic primitives by considering an AES-NI-based AES implementation as a case study, as considered previously in the \emph{Platypus} attack~\cite{Lipp2020Platypus}. 
While the \emph{Platypus} attack relies on power consumption information directly exposed to privileged software through a telemetry interface, a frequency throttling side-channel {\em converts} the power differences to power limit-induced execution time differences, which are easily accessible by malicious software.
It is to be further noted that the \emph{Platypus} attack and the proposed frequency throttling side-channel attack are two different attacks that exploit the same information leakage source, namely data-correlated power consumption resulting from inherent CMOS circuit properties.
Also, both attacks require neither physical access to the target platform nor any additional high-precision power measurement setup as required in {\em traditional} side-channel attacks.
The results of our experimental evaluations reveal the effectiveness of such an attack in successfully recovering the secret AES key, by performing statistical analysis on the collected timing side-channel traces.
In addition, we also discuss potential mitigation options to thwart throttling side-channel attacks, and their pros and cons. The main contributions of this paper can be summarized as follows:
\begin{itemize}[leftmargin=*]
    \item Presentation and comprehensive explanation of a new type of side-channel attack which exploits the workload-dependent CPU frequency adjustments performed by the power management architecture of modern processors
    \item Detailed experimental evaluation of the frequency throttling side-channel attack, showing extraction of cryptographic secrets (e.g., AES key) with different reactive limits, attack models, and systems
    \item Enumerating necessary conditions of a frequency throttling side-channel attack and providing discussion of mitigation options thwarting each of the necessary conditions
\end{itemize}

\noindent The rest of the paper is organized as follows: 
Section \ref{sec:back} presents background information of power management algorithms and related side-channel attacks.
Section \ref{sec:throttling} presents detailed reasoning behind the frequency throttling side-channel leakage and provides an overview of threat model and attack methodology
Section \ref{sec:attack} showcases an attack using the throttling side-channel against an AES-NI-based cryptographic implementation in various scenarios
Section \ref{sec:mitigation} discusses different options to mitigate a frequency throttling side-channel.
Section \ref{sec:related} discusses related work.
Section \ref{sec:conclusion} concludes the paper.

\section{Background} \label{sec:back}
\subsection{Side-Channel Analysis and Mitigations}
\subsubsection{Power Side-Channel}

Power side-channel analysis attacks exploit the fact that the dynamic power consumption $P_{dyn}$ of a digital CMOS-based circuit is data-dependent in nature ~\cite{kocher2011introduction}, as evident in the following equation:
\begin{equation}
\label{eq:P_dyn}
P_{dyn} = \alpha \cdot C \cdot V_{DD}^2 \cdot f
\end{equation}
where, $\alpha$, $C$, $V_{DD}$, and $f$ represent switching activity factor, load capacitance, supply voltage, and clock frequency, respectively.
The main objective of a power side-channel analysis attack is to retrieve a targeted secret by analyzing the data-dependent power consumption of a cryptographic implementation during a selected time window.
Traditional physical side-channel analysis techniques such as Correlation Power Analysis (CPA) perform statistical analysis on a large number of side-channel traces which are collected by varying the input data~\cite{mangard2008power}.
In a typical CPA attack, the attacker correlates the actual power consumption values $P_{a}$ from the collected power traces with the corresponding hypothetical power leakage values $P_{h}$ (as calculated using standard Hamming weight or Hamming distance power models). This is done for different key guesses using the following measure:
\begin{equation}
\label{eq:cpa}
\rho^{k} = \frac{cov(P_{a}, P_{h}^{k})}{\sigma_{P_{a}}\sigma_{P_{h}^{k}}}
\end{equation}
where, $\rho^{k}$, $\sigma_{P_{a}}$, and $\sigma_{P_{h}^{k}}$, represent the Pearson's correlation coefficient for key guess $k$, the standard deviation of the actual power values $P_{a}$, and the  standard deviation of the hypothetical power values $P_{h}^{k}$ for key guess $k$, respectively.

\subsubsection{Telemetry Side-Channel}
Software-accessible telemetry side-channel attack is an emerging topic discussed in recent works ~\cite{Lipp2020Platypus, liu2021methodology}. Unlike the traditional physical power side-channel attacks, telemetry side-channel attacks can be conducted remotely by malicious software, exploiting telemetry reading provided by the underlying hardware. For example, \emph{Platypus} \cite{Lipp2020Platypus} utilizes RAPL energy counter as the side-channel to deduce secret information such as AES keys from Intel\textsuperscript{\textregistered}~Software Guard Extensions Enclave (Intel\textsuperscript{\textregistered}~SGX Enclave). The variation in power consumption values due to processing of different data as measured by existing CPU telemetry interfaces (low sampling rates) is much less pronounced when compared to the variations as captured by traditional physical side-channel setups (high sampling rates). 
In addition, measurement inaccuracy and noise in the telemetry readings further reduce the signal-to-noise ratio (SNR) of the collected side-channel traces.
However, in spite of these challenges, it has been demonstrated in~\cite{Lipp2020Platypus} that the CPA attack is powerful enough to distinguish minute secret-dependent biases in the telemetry readings and thus, can successfully recover the underlying secret data.
In this work, we also use a variation of CPA attack (details in section~\ref{ssec:analysis_phase}) to perform statistical analysis on collected telemetry traces.

A number of possible countermeasures to thwart power telemetry attacks were listed in~\cite{Lipp2020Platypus}. These mitigation strategies include restricting user-space access (Ring-3 privilege) to \emph{powercap} driver in Linux and limiting the measurement resolution of telemetry interfaces.
In response to the \emph{Platypus} attack, the Linux \emph{powercap} driver has been updated to restrict unprivileged access to RAPL interface~\cite{Intel-RAPL-Patch, AMD-RAPL-Patch}.
In addition, Intel issued a mitigation adding white noise to reported RAPL interface readings, which is enforced when Intel SGX is enabled and can be enabled by software via a software switch~\cite{Intel-RAPL-Fuzz}. These approaches effectively thwart the CPA attack leveraging software-accessible power telemetry data.

\subsubsection{Traditional Timing Side-Channel}
\label{sssec:traditional_existing_mitigations}

\noindent In {\em traditional} timing side-channel analysis, an attacker exploits differences in execution cycles of victim code to deduce the targeted secret.
Such timing differences can arise due to data-dependent execution cycles of the victim process or mutual access of shared system resources (e.g., cache lines, branch predictors, etc.) by the victim and attacker processes.

Several countermeasures have been proposed in related literature, which use constant-cycle coding principles to address the issue of {\em traditional} timing side-channel leakage ~\cite{Intel-coding-guideline}. A short summary of such coding principles is as follows:

\begin{itemize}[leftmargin=*]
    \item Ensure code processes secret data consistently (i.e., requires same number of clock cycles irrespective of secret data values).
    \item Ensure secret data values (or values derived from secret data) do not affect the sequence of instructions executed due to a conditional branch or an indirect branch target in the code.
    \item Ensure memory access patterns (or the data size of load/store operations) are invariant with respect to secret data.
\end{itemize}
\noindent Existing work largely assumes that traditional timing side-channel leakage can be mitigated by applying these principles to the secret data-dependent portions of the code.

\subsection{CPU Power Management}
We introduce how modern CPU power management algorithms control processors' \emph{power performance states}.

\subsubsection{Processor Performance States} Intel processors implement performance states (referred to as $P$-$States$, defined per ACPI~\cite{acpi00specification}), by realizing a DVFS mechanism for optimizing power consumption. Such $P$-$States$ correspond to different $voltage$-$frequency$ pairs, which can be {\bf proactively} controlled either by the operating system (using {\em SpeedStep}~\cite{Intel-SpeedStep}) or by the hardware (using {\em Speed Shift}~\cite{Intel-Speed-Shift}). As per convention, the highest CPU $P$-$State$ is referred to as P$0$, and it corresponds to the highest achievable operating frequency, as determined during manufacturing, enabling the processor to enter the so-called {\em turbo mode}. AMD processors also implement various schemes to optimize energy efficiency, including ACPI $P$-$States$ \cite{energy-efficiency-amd-zen2}. The platform provides $P$-$States$ limits, status and voltage, frequency definitions through MSRs \cite{Open-Source-Register-Reference-AMD-17H}.   

During execution of a workload, if any system-specific limit (electrical or thermal) is violated, the processor $P$-$State$ is {\bf reactively} controlled by the power management algorithm.
The remainder of the paper focuses on reactive limits, which induce the frequency throttling side-channel. 
Depending on the criticality of the limit being hit, reactive control of $P$-$States$ may be performed with various response times (in the order of a few ms to tens of seconds) to promptly bring the system back to safe operating conditions.

\subsubsection{$P$-$State$ Control under Reactive Limits}
\label{ssec:change_p_state}
The power management algorithm of a CPU periodically calculates different running averages of electrical parameters (e.g., power, current, etc.) of windows of pre-specified lengths. \emph{Power budget} is then computed as the difference between the running averages and the respective reactive limit values. Based on the power budget, the power management algorithm $PL\_ALG(\cdot)$ computes the new $P$-$State$ limit $f_{max}$, which is the highest possible CPU operating frequency that satisfies all the reactive limits of the system. An overview of the new $f_{max}$ selection process by the controller is presented in Algorithm~\ref{algo:new_p_state}. 

When none of the reactive limits are hit, all power budgets remain positive and CPU operating frequency is not capped by $f_{max}$ (or, in other words, $f_{max}$ is higher than the highest turbo frequency). If any of these calculated running averages exceed a specific reactive limit (e.g., power limit $PL$), the power management algorithm will trigger CPU throttling activity and reduce $f_{max}$. In such cases, in order to maximize performance while satisfying the reactive limits, the processor may run at the frequency limit $f_{max}$, as governed by its feedback control mechanisms \cite{aastrom2010feedback}.

\begin{algorithm}
\caption{Determination of new $P$-$State$ limit $f_{max}$}
\label{algo:new_p_state}
\begin{algorithmic}[1]
\Require{\\
    (i) System reactive limit $PL_i$ with running average time window $\tau_i, i\in[1,N]$ \\
    (ii) Polling interval $T$\\
    (iii) Current $P$-$State$ limit $f_{max}$\\
    (iv) Power management control algorithm $PL\_ALG(\cdot)$
    }
\Ensure New $P$-$State$ limit $f_{max}$
\For{every $T$ time units}
\For{$i$ from $1$ to $N$}
\State $\overline{P_i} \gets$ \text{Calculate avg. power over} $\tau_i$
\State $\Delta \gets \overline{P_i} - {PL_i}$ \textcolor{blue}{ \ /*Available power budget*/}
\State $f_{max,i} \gets PL\_ALG(\Delta)$
\If{$f_{max} > f_{max,i}$} 
\State $f_{max} = f_{max,i}$ \textcolor{blue}{\ /*Throttling activity*/}
\EndIf
\EndFor
\EndFor
\end{algorithmic}
\end{algorithm}

\begin{figure*}[!ht]
\includegraphics[width=15cm]{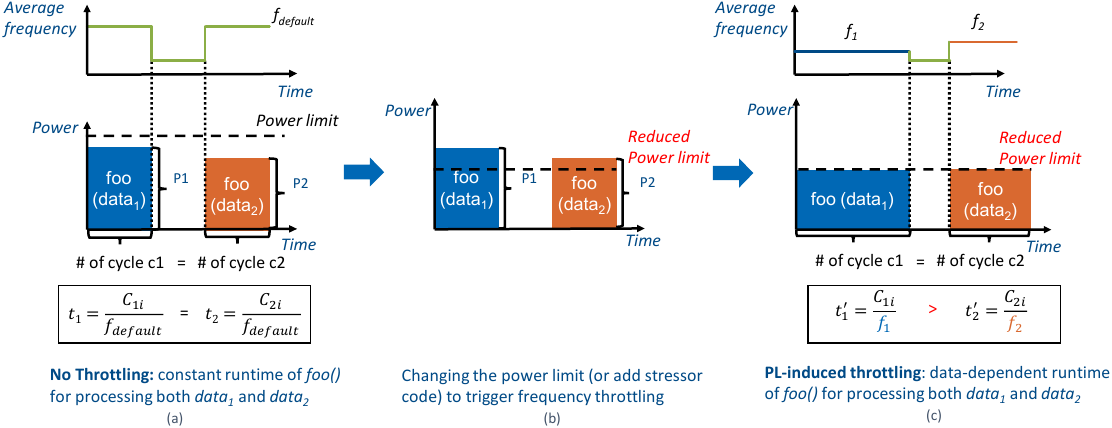}
\vspace*{-3mm}
\caption{Conversion of power side-channel to timing side-channel leakage by reactive limit-induced throttling.}
\label{fig:throttling_sc}
\end{figure*}

\subsubsection{Reactive Limits on Intel processors} 
We describe two of the \emph{reactive limits} present in several modern Intel processors~\cite{Intel-SDM}.

\begin{itemize}[leftmargin=*]
     
\item{ \bf Running Average Power Limit (RAPL):}
RAPL is a feature supported by Intel power management architecture to cap the power consumption on the system. When the configured power limit is exceeded, the CPU will be forced to run at a lower frequency to maximize performance while meeting the power limit requirement. Intel currently provides multiple power limit capabilities. The most commonly used ones are the package-level power limit 1 (PL1) and package-level power limit 2 (PL2). PL1 is used to track the long-term power consumption, so typically its limit value is set to be lower and the time window $\tau$ is longer (tens of seconds). On the other hand, PL2 is used to track the short-time power burst events, so typically the limit is set to be higher than PL1 and $\tau$ is much shorter (several milliseconds). Ring 0 software can configure the running average window $\tau$ and the power limit of each capability through interfaces such as MSRs (e.g., \emph{MSR\_PKG\_POWER\_LIMIT} for package-level power limits). 

\item{\bf Voltage Regulator Thermal Design Current Limit (VR-TDC):}
VR-TDC is a power management feature supported by Intel power management architecture. It is a current limit specified in Amperes, maintained in order to satisfy VR electrical constraints. Generally, the algorithm monitors the running average current in Amperes by reading the VR current sensor during the configured time window. If the limit is hit, the processor will engage its frequency throttling to reduce its frequency, in order to ensure current remains within the limit and budget.

\end{itemize}

\subsubsection{Reactive Limits on AMD processors}\label{sssec:reactive_limits_amd}
Precision Boost Overdrive (PBO) is a feature available on AMD Ryzen to overclock processors to achieve more performance by controlling power/thermal limits \cite{amd-pbo}. Some of the reactive limits are described as follows:

\begin{itemize}[leftmargin=*]
    \item Package Power Tracking (PPT): analogous to PL1 on Intel processors, this limit caps the total power capacity of the processor socket in Watts \cite{amd-ryzen-master}. 
    \item Package Power Tracking Fast (PPT Fast): analogous to PL2 on Intel processors, this limit is a PPT limit with faster response time (shorter $\tau$). 
    \item Thermal Design Current (TDC): analogous to VR-TDC on Intel processors, this limit caps the total current capacity in Amperes at the thermal throttling limit of the processor \cite{amd-ryzen-master}.
\end{itemize}

The aforementioned reactive limits can be configured from system software via System Management Unit (SMU) mailbox interface with support from a kernel driver \cite{Ryzen_SMU, Ryzen_monitor}. The SMU is a subcomponent of the AMD processor that is responsible for a variety of system and power management tasks during boot and runtime \cite{kernel-developer-guide-AMD-15H}. The command to configure the reactive limits is described in \cite{Ryzen_SMU}. It specifies the SMU mailbox ID and power/thermal limit value to set the reactive limit \cite{Ryzen_SMU}. The SMU mailbox IDs for the reactive limits on different AMD platform are described in \cite{ZenStates-Core}.

\section{Frequency Throttling Side-Channel} \label{sec:throttling}

In this work, we demonstrate how an attacker can leverage reactive limit triggered frequency throttling to create a novel source of timing side-channel leakage during workload execution. We show that mere application of constant-cycle coding principles is insufficient to thwart timing side-channel attacks, as system clock frequency may vary during the code execution phase and may be data-dependent to leak information.

\subsection{Attack Primitives}\label{sssec:source_throttling_sc}
In modern processors, a major source of throttling side-channel information leakage is related to the workload-dependent reactive control of $P$-$States$ (line $11$ of Algorithm~\ref{algo:new_p_state}).
Let us consider the example presented in Fig.~\ref{fig:throttling_sc} to understand the underlying implementation details leading to such throttling side-channel leakage.
Suppose that the workload under execution is a constant-cycle implementation of a function, {\em foo (arg data)}, with input argument {\em data}. As noted in section~\ref{sssec:traditional_existing_mitigations}, such a constant-cycle implementation has no traditional timing side-channel leakage. On the other hand, the power consumption of the workload due to processing different data inputs ($data_1$ and $data_2$) might vary due to the differences in internal data-dependent computations of the {\em foo} function.
Without loss of generality, let us assume that the processing of $data_1$ consumes higher power ($P_1$) compared to that of $data_2$ ($P_2$) (i.e.,  $P_1>P_2$). 
As illustrated in Fig.~\ref{fig:throttling_sc}(a), if both $P_1$ and $P_2$ are below all system-defined reactive limits, there is no throttling activity and system frequency $f_{default}$ remains the same irrespective of data consumed by {\em foo}. Therefore, in this case, there is no data-dependent timing side-channel information leakage, as the execution time of the workload is independent of the inputs.

However, when power consumption reaches or crosses the system's electrical reactive limits (e.g., the limit is configured to a lower value, or a power-hungry stressor code is executed in parallel with function {\em foo}) as illustrated using Fig.~\ref{fig:throttling_sc}(b), reactive limit induced throttling activity will be triggered, resulting in a change to $P$-$States$, as shown in Fig.~\ref{fig:throttling_sc}(c).
Since $P_1>P_2$, the average throttling frequency $f_1$ for data input $data_1$ will be lower than the average throttling frequency $f_2$ for data input $data_2$ to satisfy the {\em same} reactive limit. Both of these throttling frequencies will be lower than the default system frequency prior to throttling (i.e., $f_1<f_2<f_{default}$). \emph{Crucially},
as shown in Fig.~\ref{fig:throttling_sc}(c), the execution time of {\em foo} with frequency $f_1$ is higher compared to its execution time with frequency $f_2$.
Therefore, even though {\em foo} is a constant-cycle workload implementation, its execution time becomes data-dependent due to such frequency throttling activity.
This forms a new type of side-channel information leakage source in modern processors, which we refer to as the \textbf{frequency throttling side-channel}. 

\begin{figure}[]
\includegraphics[width=\linewidth]{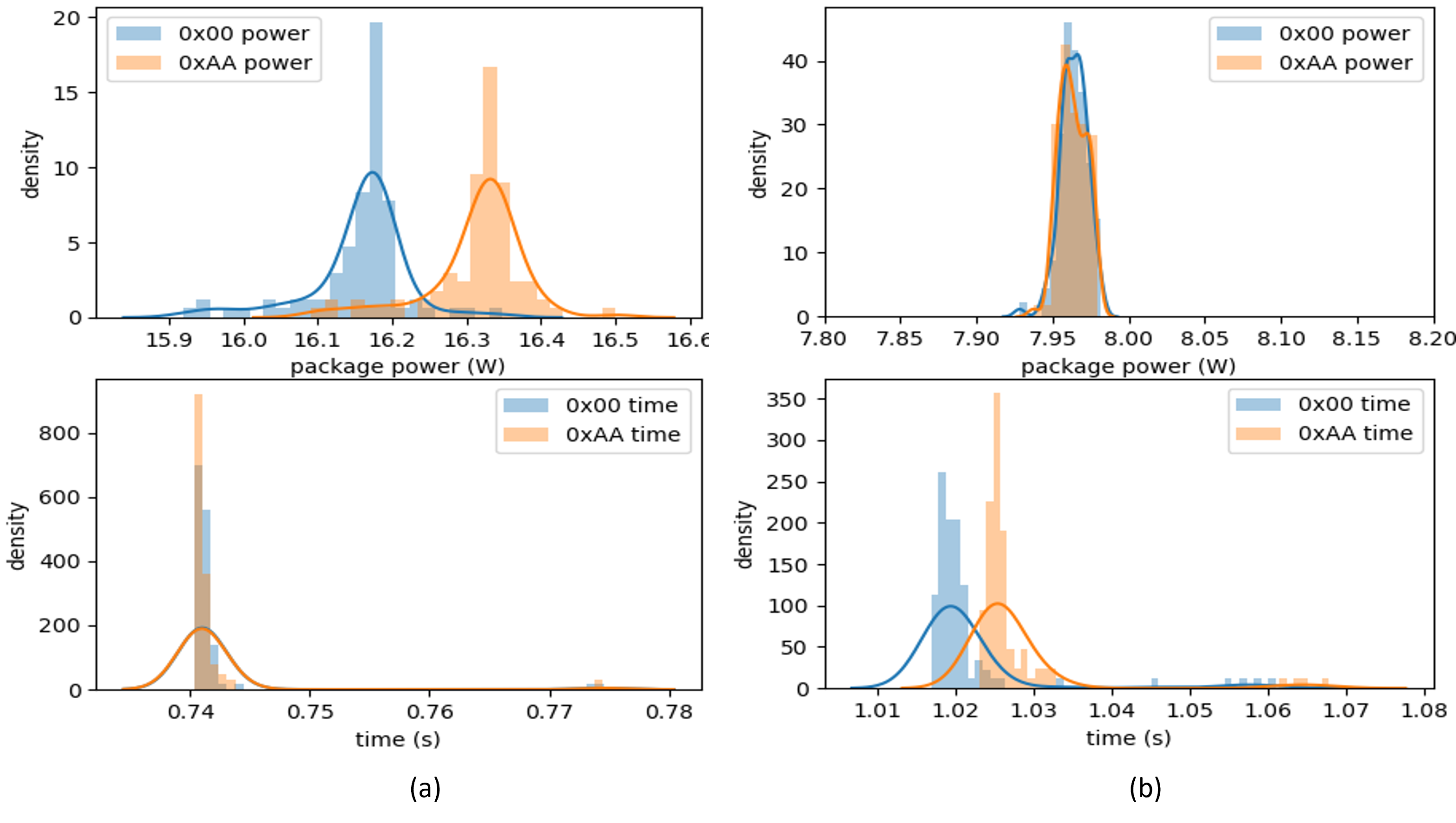}
\vspace*{-4mm}
\caption{Average power consumption and time elapsed of the same IMUL workload with different operands (a) when throttling is not triggered, and 
(b) when there is reactive limit induced throttling.}
\label{fig:imul}
\end{figure}

\begin{figure}[]
\includegraphics[width=\linewidth]{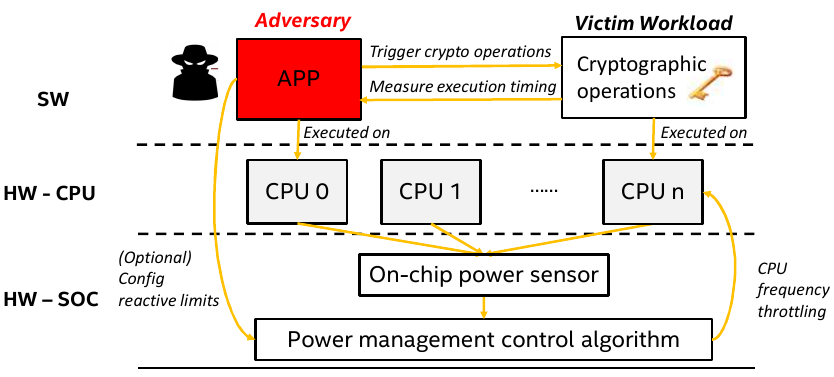}
\caption{Exploitation of frequency throttling side-channel information leakage from a victim code.}
\label{fig:threat_model}
\end{figure}

To visually appreciate the primitive, we design and run a proof-of-concept (PoC) code on an Intel E3-1230V5 system, plotted in Figure \ref{fig:imul}. The function {\em foo} we use is composed of 2.8 billion IMUL instructions, which is a cycle-constant instruction. Each IMUL instruction has one operand fixed and the other operand set to either $data_1=0x0$ or $data_2=0xAA..AA$. In the first run, we configure reactive limits to high values that will not be reached, to prevent throttling from happening, and then execute {\em foo} with $data_1$ and $data_2$, measuring aggregated package energy consumption and time elapsed for {\em foo}.  After repeating 100 times, we plot histograms of the average power consumption (calculated from energy dividing by time) and the time elapsed in Figure \ref{fig:imul} (a), respectively. As can be observed, the $0xAA$ case consumes more power than the $0x00$ case and the time it takes to execute {\em foo} for the two operands is identical. The second run duplicates the first run, except for reducing Power Limit 1 (PL1) to 8W and corresponding $PL1$ $\tau$ to 1s. With this setting, PL1 is hit, and frequency throttling is triggered. As can be seen from Figure \ref{fig:imul} (b), power consumption distributions with different data become indistinguishable and both are capped at 8W, which is the power limit. Critically, {\em foo} now takes a longer time to execute with $0xAA$ compared to $0x00$, while both are slower than the case without throttling. The results confirm that reactive limit induced throttling converts a power side-channel to a timing side-channel. Of note, Intel's Platypus mitigation can hide the information leakage exhibited in the power domain, but is not designed to mitigate the frequency throttling side-channel. Therefore, the frequency throttling side-channel is observable even if such mitigation is enabled. 

\subsection{Overview of Frequency Throttling Side-Channel Attack}

\subsubsection{Threat Model}
\label{sssec:threat_model}
We assume the following attack scenarios for an adversary exploiting a frequency throttling side-channel against a victim workload across security boundaries.

\begin{itemize}[leftmargin=*]
    \item {\bf Attack Scenario 1:} The attacker is a {\bf privileged software attacker}, such as kernel-space software or a hypervisor. The victim workload is executed inside a Trusted Execution Environment (TEE), such as Intel SGX~\cite{SGX, costan2016intel} or AMD SEV~\cite{kaplan2016amd, SEVerity}.
    Such TEEs help protect secret information of the victim application from direct access by even privileged software. In order to enable higher throttling activity during the execution of code residing in a TEE, a privileged attacker may alter reactive limit configurations. Discussion of this scenario is provided in section \ref{ssec:intel_sgx}.

    \item {\bf Attack Scenario 2:} The attacker is a {\bf user-space attacker} with Ring 3 privilege, and the victim is another application or the kernel. Unlike the previous scenario, the attacker does not have the privilege to alter the values or $\tau$ of the reactive limits. A Ring 3 attacker instead may execute a stressor code in parallel to the victim code to boost the system power consumption beyond the default limits, such that throttling activity is triggered. Note that the long $\tau$ and additional noise introduced by the stressor code will make the attack harder. Discussion of this scenario is provided in section \ref{ssec:user-space}.
    
\end{itemize}

\noindent In addition, we assume that the victim code under consideration is implemented following the constant-cycle coding principles highlighted in section~\ref{sssec:traditional_existing_mitigations}.
In order to attack such a workload, an attacker utilizes the reactive limit-induced throttling (see section~\ref{sssec:source_throttling_sc}) to make the execution time of the code vary for different {\em known} data inputs. 
The main objective of the attacker (for both of the above attack scenarios) is to deduce the targeted secret information of victim code by correlating the secret data-dependent computations with the collected execution time of the code for different data inputs.

\subsubsection{Attack Methodology}
\label{sssec:attack_method}
Fig.~\ref{fig:threat_model} presents an overview of the throttling side-channel attack against a constant-cycle implementation of victim workload. Such an attack consists of the following three phases:

\begin{itemize}[leftmargin=*]
\item {\bf Configuration Phase:} The attacker profiles a victim-like workload with a {\em known} asset to either configure the related power management settings (Ring 0 attacker) or identify a suitable stressor code (Ring 3 attacker) such that victim workload execution will result in triggering frequency throttling.  
\item {\bf Online Phase:} The attacker provides different data inputs to the victim code, which in turn performs one or more {\em secret} asset-dependent computations. The attacker also runs a monitor program in parallel to measure the execution time of the victim code with various data inputs.
\item {\bf Analysis Phase:} The attacker applies analysis methods (such as CPA) on the collected execution time values, to deduce the targeted {\em secret} asset of the victim code.
\end{itemize}
\section{Case study: Attack against AES encryption} \label{sec:attack}

In this section, we consider a victim workload comprising an AES-NI-based implementation of AES-128 as a case study to illustrate the proposed throttling side-channel attack. 
In order to assess the side-channel leakage due to throttling activity, we use statistical analysis methods derived from {\em Test Vector Leakage Assessment (TVLA)}~\cite{Goodwill11,Standaert2017HowT} and
correlation-based analysis.

\subsection{Victim Workload}\label{sec:victim}

\subsubsection{AES-128 Algorithm}
AES is a block cipher established by NIST in 2001~\cite{FIPS197}. The algorithm encrypts a fixed size plaintext block of 128-bit using key-size of 128, 192, or 256 bits and outputs a 128-bit ciphertext. In this work, we consider the AES-128 primitive to demonstrate frequency throttling side-channel attack.
\subsubsection{AES-NI-Based AES Implementation}
AES-NI is an instruction set which improves the AES implementation by accelerating its complex performance-intensive steps using dedicated hardware~\cite{gueron2010intel}. AES-NI instructions also provide improved security against side-channel attacks due to their constant-cycle implementations. AES-NI instructions are supported by x86 processors of major vendors, including Intel and AMD. To perform encryption, first individual round keys are derived using the \texttt{KeyExpansion} procedure~\cite{gueron2010intel}. This is followed by the \texttt{Initial AddRoundKey} operation, which computes the bitwise XOR between the plaintext and the initial round key. \texttt{AESENC} instruction is then used to perform a single round of encryption with round state and round key as the input operands. The \texttt{AESENCLAST} instruction performs the last round of encryption and returns the ciphertext.

\subsection{Attack Methodology Details}\label{ssec:attack_method}
As outlined in section~\ref{sssec:attack_method}, the frequency throttling side-channel attack against a victim workload comprises three distinct phases: the configuration, the online, and the analysis phases.
Next, we present the details of each of these phases as adopted in our case study, using power limit as an example.

\subsubsection{Configuration phase}\label{ssec:offline}
During the configuration phase, the privileged software attacker first profiles the AES-NI-based victim workload to estimate $P$, which is the power consumption of the victim workload without frequency throttling. Then, in order to trigger throttling activity, a privileged software attacker may adjust the system's power limit value $PL$ such that $P>PL$.
As per Algorithm~\ref{algo:new_p_state}, such an adjustment will lead to a change in $P$-$State$ that satisfies the available power budget.
In order to ensure that there is data-dependent frequency throttling activity, the value of the reactive limit $PL$ should be carefully adjusted$:$ setting the $PL$ value too high will not satisfy the requirement of $P>PL$ whereas setting the $PL$ value too low will cause the system to execute in a constant low frequency. For a user-space attacker who does not have the privilege to configure $PL$, an alternative approach is to run a stressor workload to boost power consumption. The details will be discussed in section \ref{ssec:user-space}.
\subsubsection{Online phase}\label{ssec:online}
During the online phase, the attacker inputs plaintext to the victim workload and obtains the corresponding ciphertext as output. Note that due to PL-induced throttling activity, different plaintexts will result in different processing times which correspond to the side-channel traces.
During the $i^{th}$ trace collection, the attacker first records the starting time stamp counter value ($T^i_1$) just before sending the $i^{th}$ plaintext to the victim workload and subsequently, also records the end time stamp counter value ($T^i_2$) just after receiving the $i^{th}$ ciphertext. Then, the attacker calculates the corresponding execution time $T_\delta^{i}$ as $T^i_2-T^i_1$.
On most of the modern CPUs, incrementing of the time stamp counter is \emph{frequency-invariant} so the $T_\delta$ captures wall clock time of the victim's execution time and will not be impacted by frequency throttling~\cite {Intel-SDM,AMD-user-guide}.

\noindent{\bf Techniques to reduce Minimum Time to Disclosure:}
We define $Minimum~Time~to~Disclosure~$(MTD) as the minimum time spent to collecting enough side-channel traces to recover the secret information.
Reduction in MTD can be achieved by (i) increasing the Signal-to-Noise (SNR) of the captured traces, or
(ii) decreasing the collection time of each trace~\cite{mangard2008power}.
In our experiments, we adopted the following techniques to lower the MTD of frequency throttling side-channel attack:

\begin{itemize}[leftmargin=*]

\item Multiple instances of the victim workload were executed simultaneously across different processor cores to amplify its average power consumption, which in turn results in higher SNR of the collected traces.

\item Each instance of the victim workload was executed repeatedly $N$ times with the same input to collect one trace.
This approach not only ensured that the on-chip sensors accurately sampled the average power consumption of the victim workload during a reactive limit's $\tau$ window but also boosted the SNR of collected traces for a sufficiently large value of $N$ (which spans across multiple $\tau$ windows) due to the denoising effect of {\em averaging}~\cite{wu2020remove}.

\item Selection of a system reactive limit having the lowest possible configurable value of $\tau$ to trigger throttling activity during victim workload execution. For a chosen value of $N$, the trace collection time corresponding to a reactive limit having lower $\tau$ will be shorter compared to that corresponding to a reactive limit with higher $\tau$. Of course, such selection or adjustment of a reactive limit is only applicable to a privileged attacker.

\end{itemize}

\subsubsection{Analysis phase}
\label{ssec:analysis_phase}
We utilize two statistical techniques to analyze potential side-channel information leakage arising from throttling side-channel activity:
(i) First, in order to ascertain if different data exhibit different PL-induced throttling behavior, we apply TVLA to the corresponding timing traces.
(ii) Second, depending upon the positive outcome of the TVLA, we perform a CPA attack in order to determine if the targeted AES encryption key can be recovered by analyzing the collected timing side-channel traces.
Next, we present the details of these statistical techniques as adopted in our experiments.

\noindent{\bf TVLA:} TVLA methodology utilizes t-scores generated from Welch’s t-test to assess potential side-channel leakage in cryptographic implementations \cite{gilbert2011testing}. Welch's t-test defines a statistical measure, {\em t-score}, as shown in equation (\ref{eq:t-score}) to compare two datasets A and B

\begin{equation}
\label{eq:t-score}
    \text{$t$-$score$} = \frac{\mu_1 - \mu_2}{\sqrt{s_1^2/N_1+s_2^2/N_2}} 
\end{equation}

\noindent where, $\mu_1$, $\mu_2$ are sample means, $s_1^{2}$, $s_2^{2}$  are sample variances, and $N_1$,  $N_2$ are the number of samples in datasets A and B, respectively.  
A |$t-score$| $>$ 4.5 rejects the null hypothesis with 99.999\% confidence, indicating datasets A and B are statistically distinguishable \cite{tobias-leakage-assessment}.

In this case study, we assess the throttling side-channel leakage of an AES-NI-based AES implementation victim workload using the above-mentioned TVLA methodology. We collect the timing traces corresponding to encryption of three different sets of plaintexts: $All\_one$, $All\_zero$ and $Random$. We apply TVLA test on timing traces for all possible pairs of chosen plaintexts.  

\noindent{\bf CPA Attack:}
We also apply CPA attack to recover the secret key from the AES-NI based victim workload by collecting timing traces $T_{\delta}$ corresponding to different randomly generated plaintexts.
Then, we utilized the following equation to calculate the Pearson's correlation coefficient metric $\gamma^{k}$ for different key guess $k$.
\begin{equation}
\label{eq:CPA}
\gamma^{k} = \frac{cov(T_{\delta}, T_{h}^{k})}{\sigma_{T_{\delta}}\sigma_{T_{h}^{k}}}
\end{equation}
\noindent where, $\sigma_{T_{\delta}}$ and $\sigma_{T_{h}^{k}}$ represent the standard deviations of the actual PL-induced execution time traces $T_{\delta}$ and the hypothetical execution time estimations $T_{h}^{k}$, respectively.
Note that similar to power estimates, such execution time estimates can also be obtained using standard Hamming weight (HW) or Hamming distance (HD) models. 
This is because the PL-induced execution time variations of the victim workload are proportional to the corresponding power consumption variations. 

The attacker may choose either the initial round (targeting initial \texttt{AddRoundKey} round key $k_0$) or the last round (targeting last round key $k_{10}$) of AES implementation as \emph{point of attack}. Intermediate AES rounds are typically not considered as attack points because in those cases the corresponding hypothetical execution time values become a function of multiple round keys, thus substantially increasing attack complexity. Selection of hypothetical execution time model ($T_{h}^{k}$) is based on the point of attack \cite{cryptoeprint:2016/700, HW-HD-power-model}:
\begin{itemize}[leftmargin=*]
    \item \textbf{Round0-HW}: To recover the initial round key $k_0$, the hypothetical execution time values can be modeled using HW of the initial \texttt{AddRoundKey} output.
    \item \textbf{Round10-HW}:To recover the last round key $k_{10}$, the hypothetical execution time values can be modeled using HW of last round input, derived using ciphertext and hypothesis of $k_{10}$.
    \item \textbf{Round10-HD}: Also, the attacker can model the hypothetical execution time values using HD between last round input and ciphertext, derived using ciphertext and hypothesis of $k_{10}$.
    \item \textbf{Round10-HW+HD}: We also considered a leakage model comprising the sum of \textbf{Round10-HW} and \textbf{Round10-HD} models, derived using ciphertext and hypothesis of $k_{10}$.
\end{itemize}

We use \emph{Guessing Entropy} (GE) to evaluate the success of a CPA attack, based on the metric from \cite{RivainSAC08}. For byte $i$ of the round key, all possible key guesses are sorted in descending order of their correlation coefficient, to obtain the rank of the correct key byte. For all bytes of the guessed key, the ranks are summed up (logarithmically) to get GE, as shown in Equation (\ref{eq:GE}).

\begin{equation}
\label{eq:GE}
    \text{GE} = \sum_{i=1}^{16}log_{2}[rank(k_{i}^{correct\_key})]
\end{equation}

Lower GE indicates lower average ranks across all key bytes and, most of the time, means more key bytes are recovered successfully.
\textbf{A value of GE=0 implies all key bytes have been recovered}. In practice, a successful CPA attack should result in a GE value lower than a pre-defined threshold which allows the recovery of the key bytes with reasonable computational complexity.

\subsection{Evaluation}
\label{ssec:eval}

\begin{table*}[]
\centering
\caption{Information and configurations of the systems under test.}
\label{tab:system}
\vspace*{-3mm}
\resizebox{15cm}{!}{%
\begin{tabular}{c|ccccccc}
      Processor Number            & \# of physical cores& Max Turbo frequency & SMT                                                     & PL2/PPT Fast limit value & PL2 $\tau$ \\ \hline

Intel E3-1230V5        & 4                     & 3.8 GHz             & disabled                                                & 10W               & 2ms        \\ \hline
Intel i7-1185G7      & 2                    & 4.1 GHz             & disabled & 8W                & 2ms        \\ \hline

Intel Xeon Gold 6326  & 8/socket              & 3.5 GHz             & disabled & 50W               & 2ms \\ \hline

AMD Ryzen 5 5600G  & 6              & 3.2 GHz             & disabled & 15W               & -  \\ \hline

\end{tabular}
}
\end{table*}

\begin{table*}[]
\centering
\caption{Pairwise t-score (absolute value) among $All\_zero$, $All\_one$, and $Random$ traces. T-score greater than 4.5 are marked in \textbf{bold} and indicates the set of data are statistically distinguishable.}
\label{tab:TVLA}
\vspace*{-3mm}
\resizebox{15cm}{!}{%
\begin{tabular}{l|ccc|ccc|ccc|ccc}
             & \multicolumn{3}{c|}{Intel E3-1230V5}                                                                        & \multicolumn{3}{c|}{Intel i7-1185G7}                                                            & \multicolumn{3}{c}{Intel Xeon Gold 6326}                     
                        \\
             & \multicolumn{1}{c}{All\_zero\_2} & \multicolumn{1}{c}{All\_one\_2} & \multicolumn{1}{c|}{Random\_2} & All\_zero\_2                 & All\_one\_2                  & Random\_2                   & All\_zero\_2                 & All\_one\_2                  & Random\_2                                   \\
             \hline
All\_zero\_1 & {\color[HTML]{333333} 1.02}      & {\color[HTML]{FE0000} \textbf{41.02}}    & {\color[HTML]{FE0000} \textbf{26.20}}   & {\color[HTML]{333333} 0.23}  & {\color[HTML]{FE0000} \textbf{10.57}} & {\color[HTML]{FE0000} \textbf{5.32}} & {\color[HTML]{333333} 0.45}  & {\color[HTML]{FE0000} \textbf{19.82}} & {\color[HTML]{FE0000} \textbf{12.51}} \\

All\_one\_1  & {\color[HTML]{FE0000} \textbf{38.42}}     & {\color[HTML]{333333} 2.38}     & {\color[HTML]{FE0000} \textbf{11.79}}   & {\color[HTML]{FE0000} \textbf{12.58}} & {\color[HTML]{333333} 3.02}  & {\color[HTML]{FE0000} \textbf{8.08}} & {\color[HTML]{FE0000} \textbf{19.74}} & {\color[HTML]{333333} 0.53}  & {\color[HTML]{FE0000} \textbf{7.57}}  \\
Random\_1    & {\color[HTML]{FE0000} \textbf{24.07}}     & {\color[HTML]{FE0000} \textbf{14.28}}    & {\color[HTML]{333333} 2.36}   & {\color[HTML]{212529} 1.32}  & {\color[HTML]{FE0000} \textbf{9.23}}  & {\color[HTML]{333333} 3.84} & {\color[HTML]{FE0000} \textbf{11.27}} & {\color[HTML]{FE0000} \textbf{7.72}}  & {\color[HTML]{333333} 0.68} 
\end{tabular}%
}
\end{table*}

In this section, we implement a Proof-of-Concept (PoC) code to demonstrate secret key recovery from an AES-NI based victim workload as outlined in section \ref{sec:victim}. Our experiments mainly focus on Intel systems, where we consider both power limit induced and current limit induced frequency throttling activity during trace collection. We also demonstrate cross-platform applicability of the attack through experimental evaluation on an AMD processor.

\subsubsection{Experimental Setup}
In our PoC, the AES encryption workload is repeated (with the same plaintext and key for encryption) for many iterations. On the same thread, the measurement code measures the aggregated execution time of the victim using \emph{time stamp counter} and logs it a single trace. The number of iterations is calibrated on each system such that every trace spans approximately 45ms (i.e., $T_\delta$=45ms). To boost SNR, multiple instances of the same victim workload are executed in parallel on the other cores.  

For experimental evaluations, we considered three Intel systems: E3-1230V5, i7-1185G7, and Xeon Gold 6326 and one AMD system: Ryzen 5 5600G. Table~\ref{tab:system} lists the details of different systems along with their corresponding PL2 (or PPT Fast for AMD system) values as adjusted in the configuration phase of the attack to introduce frequency throttling activity during workload execution.
The reason behind selection of PL2 (PPT Fast for AMD system) is due to the lower default time window $\tau$, which can be configured further to a shorter window of 2ms in an Intel system.
Note that a lower $\tau$ value reduces the minimal trace length and hence, reduces the MTD requirement as discussed in section~\ref{ssec:online}.
Also, note that since Xeon Gold 6326 is a 2-socket server system with power limits being defined independently per socket, we run the victim workload on socket 0 and adjust only PL2 corresponding to socket 0.

We first report the experimental outcomes of TVLA tests to highlight potential side-channel information leakage arising from frequency throttling activity. Then, we present the results of a CPA attack to demonstrate how an attacker can successfully recover the secret key by collecting timing side-channel traces of the AES implementation. We first test the victim code outside a TEE to showcase the behavior. Experimental results targeting a victim workload inside SGX are presented thereafter in section \ref{ssec:intel_sgx}, and experimental results with a user-space attacker is presented in section \ref{ssec:user-space}.

\subsubsection{TVLA Results}
For every system under test, following the TVLA methodology described in section \ref{ssec:analysis_phase}, we collected $10,000$ timing traces of the victim workload
corresponding to the encryption of each of the plaintext sets ( $All\_one\_1$, $All\_zero\_1$ and $Random\_1$).
We repeated the trace collection process with the {\em same} plaintext sets ($All\_one\_2$, $All\_zero\_2$ and $Random\_2$) in order to ascertain that there are no false positives in the TVLA test outcomes due to issues associated with data collection (e.g., inconsistent system behaviors or settings).
Subsequently, we first removed outlier traces followed by calculation of Welch's $t$-$score$ measure using the remaining trace sets, which correspond to every possible pairs of plaintext sets.
The results of this experiment are presented in Table \ref{tab:TVLA}, where the $t$-$score$ values greater than 4.5 are marked in \textcolor{red}{{\bf bold}} to indicate that the pairs of datasets are statistically distinguishable.
It can be observed that across all the three systems, $t$-$score$ values between trace sets corresponding to different plaintext sets are higher than 4.5 (the only exception is the case $Random_1$ vs. $All\_zero\_2$ on i7-1185G7 system) while the $t$-$score$ values between trace sets corresponding to same plaintext sets are lower than 4.5.
This signifies that, due to power limit-induced throttling activity, the CPU frequency changes in a data-dependent manner, which leads even a constant-cycle victim code implementation to exhibit data-dependent runtime differences.

\begin{table}[]
\centering
\caption{Converged Guessing Entropy (GE) on different systems with different leakage models. Lower GE implies more key bytes are recovered.}
\label{tab:GC_converged}
\vspace*{-3mm}
\resizebox{\columnwidth}{!}{
\begin{tabular}{l|cccc}
              & Intel E3 & Intel i7 & Intel Xeon & AMD Ryzen  \\ \hline
Round0-HW     & 17.5    & 27.5      & 2.6 & 2.0           \\
Round10-HW    & 85.7    & 73.3      & 86.0 & 3.16          \\
Round10-HD    & 0  & 27.3      & 81.2 & 35.8          \\
Round10-HW+HD & 0  & 21.4      & 72.7 & 1.5        
\end{tabular}
}
\vspace*{-3mm}
\end{table}

\subsubsection{CPA Attack Results with Power Limit}
With the TVLA tests exhibiting positive signs of potential side-channel information leakage, we further investigated the applicability of a CPA attack on the power limit-induced timing traces, to recover the secret AES key.
On each of the Intel systems, we collected 8 million timing traces of the PoC code (with a fixed key) by providing randomly generated plaintexts.
The trace collection process took about 100 hours on average across different systems.
After the trace collection phase, we computed the median $\mu$ of the collected traces and discarded the outliers of $[0.95\mu, 1.05\mu]$.
Subsequently, we applied the CPA attack on the filtered trace dataset, targeting the round keys $k_0$ and $k_{10}$. 

\begin{figure}[t]
\includegraphics[width=7cm]{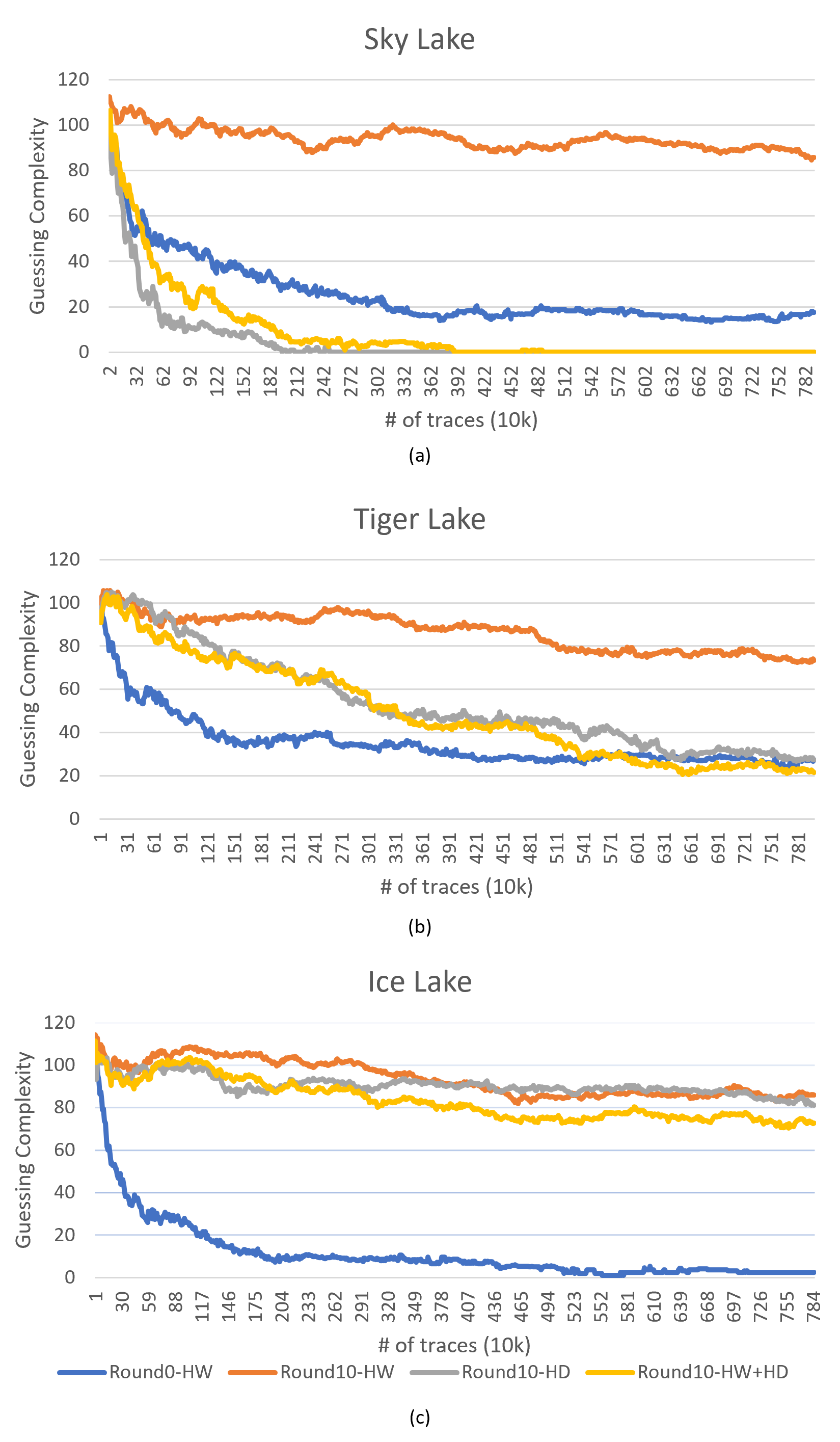}
\vspace*{-3mm}
\caption{Guessing Entropy trend with different amount of traces on (a) E3-1230V5, (b) i7-1185G7, and (c) Xeon Gold 6326 systems with power limit-induced frequency throttling}
\label{fig:cpa}
\vspace*{-2mm}
\end{figure}

In Figure~\ref{fig:cpa}, we present the GE trends (corresponding to different execution time estimate models) versus the number of timing traces considered for the CPA attack across multiple systems. Based on the data, we make the following observations:
\begin{itemize}[leftmargin=*]
    \item \textbf{The general trend} is that GE converges gradually when increasing number of traces are used for analysis. This is because a larger number of traces helps to reduce the effect of noise in the collected traces. Such GE trends highlight the fact that all the execution time estimate models considered correlate with the actual execution times of the PoC code. Also, it can be observed that in all cases, GE values start from somewhere around 112. This is because the expected value of the initial rank of correct key byte (without parsing any side-channel traces) is 128 among the 256 possible key byte guesses. Therefore, the expected value of initial GE value is $E(GE) = \sum_{i=1}^{16}log_{2}(128)=112$.
    \item \textbf{Round0-HW model} appears to be effective on all the three systems: GE converges to 17.5 on E3-1230V5, 27.5 on i7-1185G7, and 2.6 on Xeon Gold 6326. Especially, on Xeon Gold 6326, with this execution time estimate model, we successfully recovered \textbf{14 out of the 16 bytes of the correct key}. The ranks of the remaining two key bytes are 2 and 3.
    \item \textbf{Round10-HW model} converges the slowest among all the models tested. On both E3-1230V5 and Xeon Gold 6326 systems, even after analyzing with 8M traces the GE values remain above 80, signifying that the CPA attack was unsuccessful in these cases. The lowest GE value obtained was 73.3 on i7-1185G7 system.
    \item \textbf{Round10-HD model} shows distinctly different behaviors on different systems. On E3-1230V5 system, for example, GE converges to 0 (\textbf{all 16 key bytes recovered}) with less than 2M traces. Also, on i7-1185G7 system, for this model GE converges to 27.3, similar to Round0-HW model. But in the case of the Xeon Gold 6326 system, for this model the GE value reduces to only 81.2.
    \item \textbf{Round10-HW+HD model} results in consistently lower GE values compared to the Round10-HW model across all the three systems.
    However, compared to the Round10-HD model, for this model the GE value converges slower on E3-1230V5 system whereas the GE values converge faster on i7-1185G7 and Xeon Gold 6326 systems.
\end{itemize}

\noindent Note that for a given execution time estimate model, the difference in behavior of GE trends on different systems is likely due to variations in the underlying hardware micro-architecture designs and the fabrication technologies used. Table~\ref{tab:GC_converged} summarizes the outcomes of the CPA attack corresponding to different execution time estimate models across systems. 
These results demonstrate the fact that power-limit induced frequency throttling activity can be successfully leveraged by an attacker to extract secret information from cryptographic workloads.

\begin{figure}[]
\includegraphics[width=7cm]{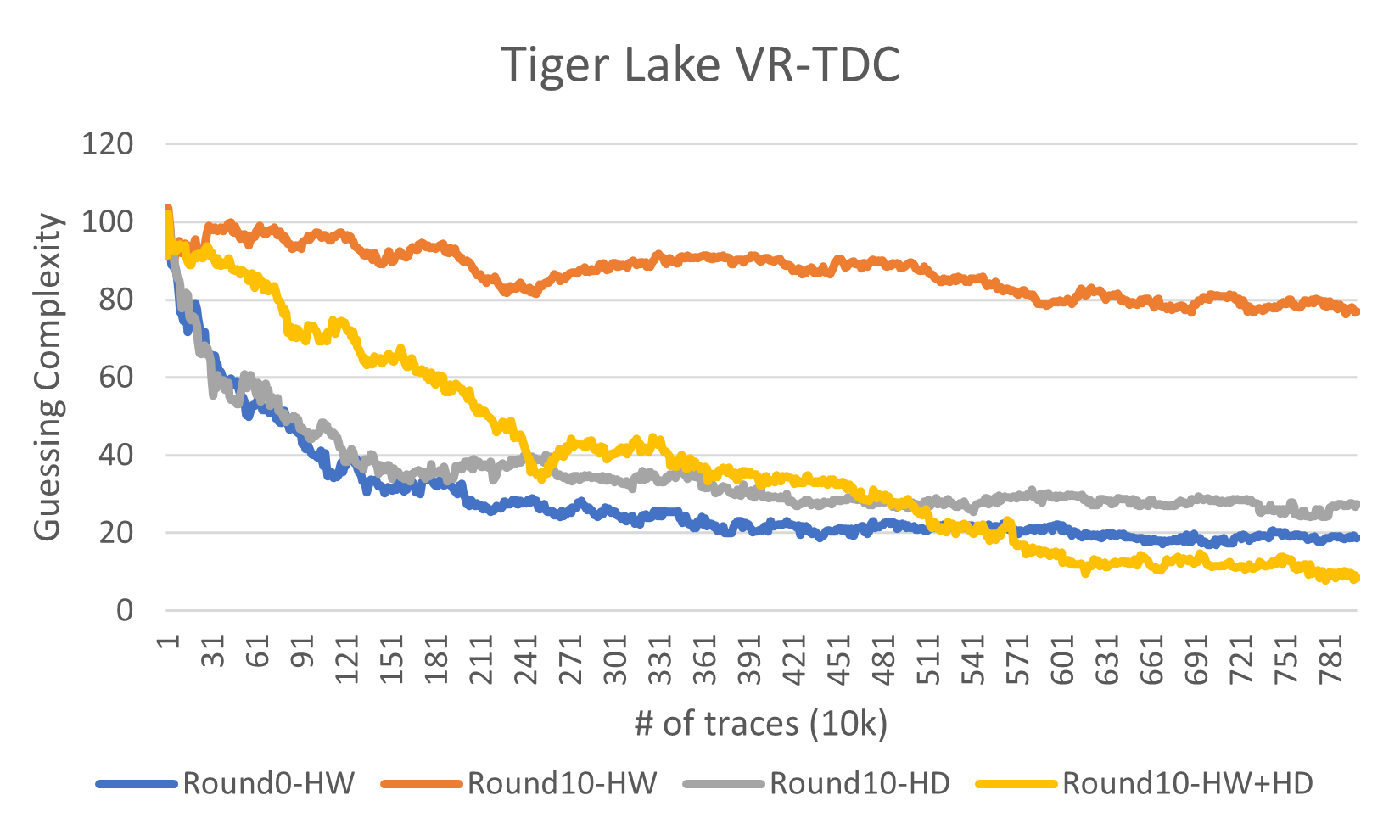}
\vspace*{-2mm}
\caption{Guessing Entropy trend with different amount of traces on i7-1185G7, with current limit-induced frequency throttling}
\label{fig:vr_tdc}
\vspace*{-4mm}
\end{figure}

\begin{figure}
\includegraphics[width=7cm]{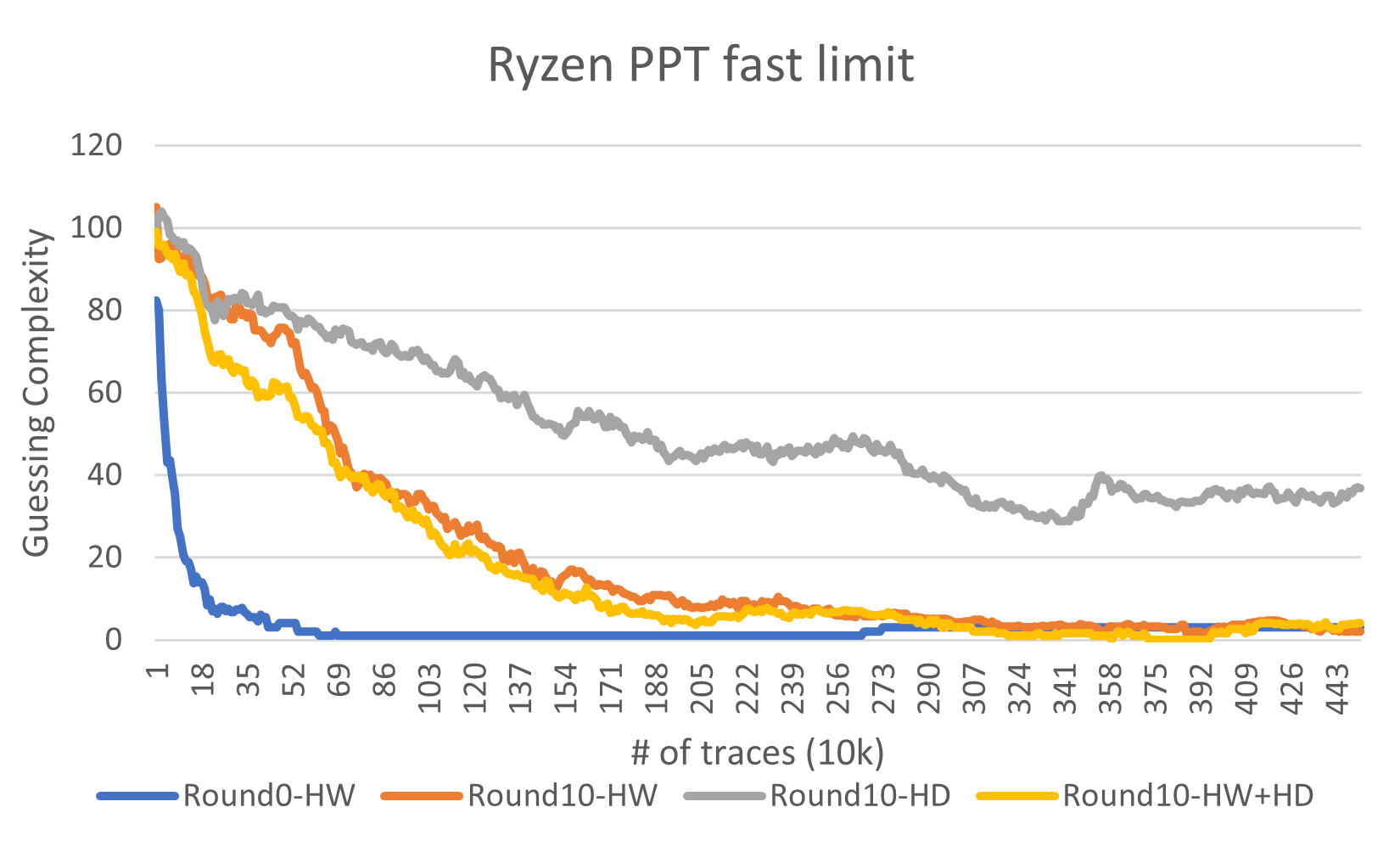}
\vspace*{-2mm}
\caption{Guessing Entropy trend with different amount of traces on an AMD Ryzen 5 5600G processor with PPT Fast limit-induced frequency throttling}
\label{fig:Ryzen}
\vspace*{-2mm}
\end{figure}

\subsubsection{CPA Attack Results with Current Limit}
We also repeated the CPA test on the i7-1185G7 system with VR-TDC limit set to 7 Amperes to trigger frequency throttling. 
All other configurations were kept the same as used for the power-limit induced frequency throttling experiments in the previous subsection.
The GE trends for different execution time estimate models are shown in Figure \ref{fig:vr_tdc}. Similar to the power-limit experiments, the GE value corresponding to the Round10-HW model converges the slowest amongst all the models tested.
The GE values for both Round0-HW and Round10-HD models converge to around 20 after analyzing with 8M traces.
The GE value corresponding to Round10-HW+HD model converges to the lowest value (around 10) for the CPA attack with current limit.
These observations confirm the fact that VR-TDC limit can also be leveraged by an attacker to mount the frequency throttling side-channel attack.

\subsubsection{CPA Attack Results on AMD Ryzen 5}

We also repeated the CPA test on an AMD Ryzen 5 5600G system with PPT Fast Limit set to 15W to trigger frequency throttling.
The victim AES workload was executed using the configuration mentioned in last row of Table~\ref{tab:system} and the number of iterations of the workload was calibrated such that every trace spans approximately 50ms (i.e., $T_{\delta}$=50ms).
We collected 4.5M traces with trace collection time of about 63 hours.

Figure \ref{fig:Ryzen} shows the trend in GE against the number of traces collected for CPA analysis across different execution timing estimate models. From the figure, we can observe that GE converges much faster with Round0-HW model as compared to other execution timing estimate models. 
The GE value converges to almost zero for approximately 1 million traces (in 16 hours) with Round-0 HW model, \textbf{recovering 14 out of the 16 secret key bytes}. A similar trend in GE was observed for other models as well; the GE values converged to (a) almost zero for approximately 3 million traces with Round10-HW and Round10-HW+HD models and (b) about 40 for approximately 4.5 million traces with Round10-HD model.
Table \ref{tab:GC_converged} summarizes the final converged GE values with different execution time estimate models on the systems we tested.

\subsection{Attacking AES-NI inside Intel\textsuperscript{\textregistered}~SGX enclave}
\label{ssec:intel_sgx}
We also evaluated an AES-NI based AES implementation from Intel’s Integrated Performance Primitives (Intel IPP)~\cite{ipp-crypto} as the victim workload, executed inside an Intel SGX enclave. The victim takes a 16-byte plaintext as the input from Enclave Call (ECALL) and encrypts it using a secret key owned by the enclave. Similar to the experiment setup in section \ref{ssec:eval}, the victim workload repeatedly encrypts the same plaintext many iterations to boost SNR: \texttt{ippsAESEncryptECB} is invoked 10000 times per trace, and each invocation encrypts 1024 16-bytes of the same plaintext. In entirety, 160 MB of data is encrypted in each \texttt{ECALL}. Ciphertext is returned to the caller of the enclave after encryption. Also, to boost SNR, we invoke multiple enclave instances, executing on multiple cores in parallel with the same plaintext as input. Only the execution time for one enclave is recorded as timing traces. 

We performed the experimental evaluation on an Intel E3-1230V5 system, with PL2 limit and $\tau$ set to 20W and 2ms, respectively. We followed TVLA test methodology as described in section \ref{ssec:analysis_phase} and collected 10000 timing traces corresponding to the encryption of each of the plaintexts ($All\_one$, $All\_zero$ and $Random$). The pairwise t-scores are presented in Table~\ref{tab:SGX_TVLA}. We observed that the t-score values between timing traces corresponding to different plaintexts are greater than 4.5, while the t-score values between timing traces corresponding to the same plaintext are lower than 4.5, indicating data-dependent leakage in the time domain, due to PL-induced throttling activity. The only exception is \emph{Random\_2} vs. \emph{All\_zero\_1}, which shows a t-score value less than 4.5. Upon comparing the TVLA results on the same system (E3-1230V5) for the victim workload outside an SGX enclave (Table~\ref{tab:TVLA}), we observe the magnitude of the t-scores (in most cases) for the victim executing inside an SGX enclave is lower. This is due to the additional noise introduced during enclave transitions, as well as IPP crypto function calls.

\begin{table}[h]
\caption{Pairwise t-score (absolute value) among $All\_zero$, $All\_one$, and $Random$ traces, test on E3-1230V5 with victim workload inside Intel SGX enclave.}
\label{tab:SGX_TVLA}
\vspace*{-3mm}
\begin{tabular}{l|ccc|}
             & \multicolumn{3}{c|}{E3-1230V5 (Intel SGX victim workload)}                                                                   \\
             & All\_zero\_2                          & All\_one\_2                           & Random\_2                             \\ \hline
All\_zero\_1 & 1.19                                  & {\color[HTML]{FE0000} \textbf{29.19}} & {\color[HTML]{FE0000} \textbf{24.31}} \\
All\_one\_1  & {\color[HTML]{FE0000} \textbf{33.12}} & 3.65                                  & 2.20  \\
Random\_1    & {\color[HTML]{FE0000} \textbf{26.47}}  & {\color[HTML]{FE0000} \textbf{6.17}} & 0.96
\end{tabular}
\end{table}

\begin{figure}[]
\includegraphics[width=7cm]{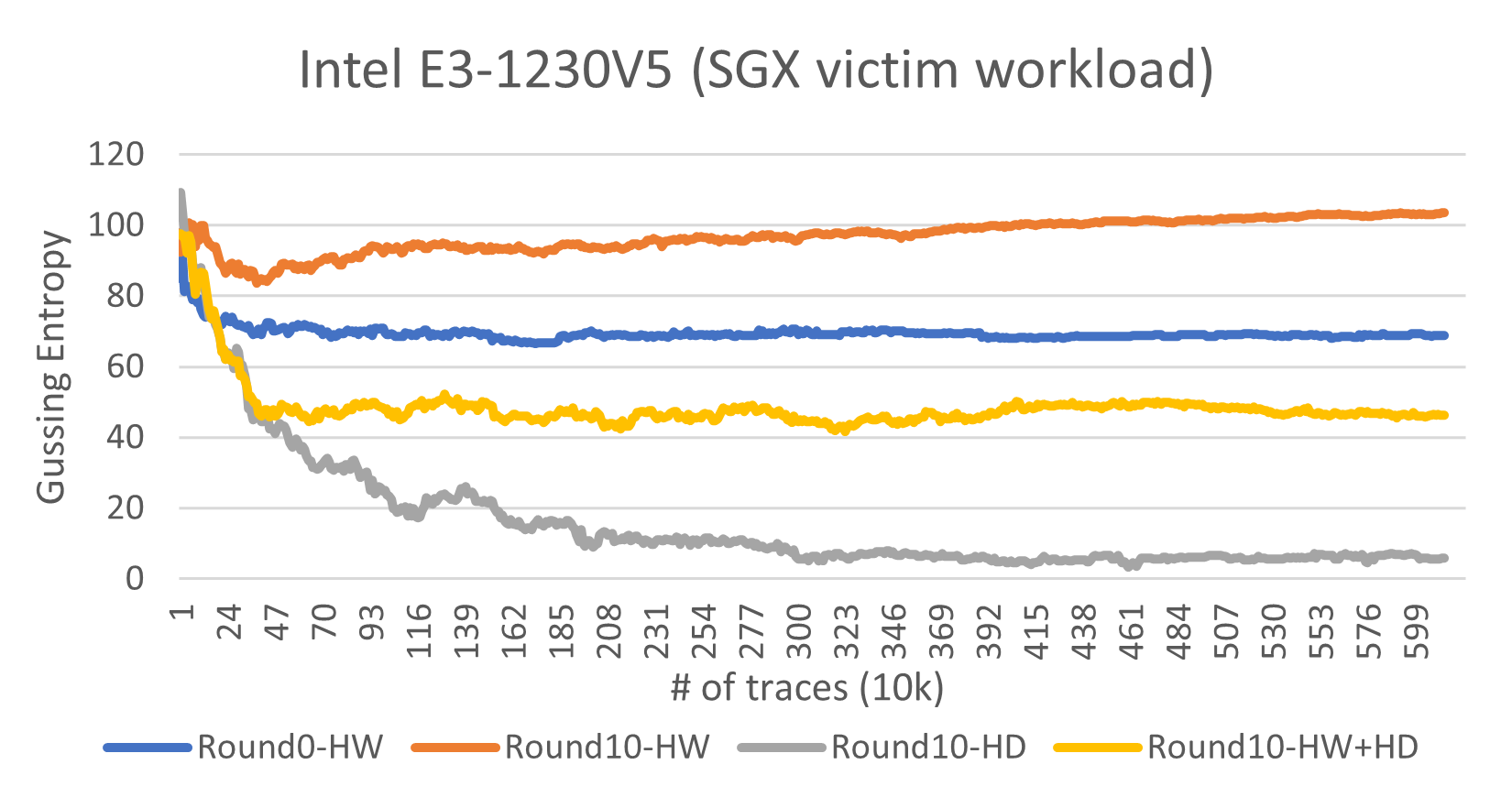}
\caption{Guessing Entropy trend with different amount of traces on E3-1230V5, with power limit-induced frequency throttling and victim in Intel\textsuperscript{\textregistered}~SGX enclave.}
\label{fig:SGX_cpa}
\end{figure}

In a CPA test, we collected 8 million traces with $T_\delta$=45ms per trace, which took about 100 hours in total. The traces are analyzed with different leakage models and the results are shown in Figure~\ref{fig:SGX_cpa}. With Round10-HD model, guessing entropy converges to 6 (\textbf{14 out of 16 key bytes recovered}) with the 8 million traces, confirming that information is observable through the throttling side-channel for the given victim. Compared to the previous CPA result on E3-1230V5 system (Figure~\ref{fig:cpa}(a)), the additional noise results in lower SNR, and hence slower guessing entropy convergence (less key bytes being revealed with the same amount of traces) for all the leakage models. For example, in the previous test on the same system (Figure \ref{fig:cpa}(a)), a full key can be recovered from a victim outside an SGX enclave with less than 2 million traces (25 hours).

\begin{table*}[]
\caption{Pairwise t-score (absolute value) among $All\_0$, $All\_1$, and $random$ traces with different PL1 settings. T-score greater than 4.5 are marked in red and indicates the set of data are statistically distinguishable.}
\label{tab:PL1-TVLA}
\vspace*{-3mm}
\resizebox{\textwidth}{!}{%
\begin{tabular}{l|lll|lll|ccc|ccc}
                                  & \multicolumn{3}{c|}{PL1=80W, w/o stressor}                                                                           & \multicolumn{3}{c|}{PL1=110W, w/o stressor}                                                         & \multicolumn{3}{c|}{PL1=110W, w/ stressor}                                                                            & \multicolumn{3}{c}{PL1=140W, w/stressor}                                                 \\
                                  & \multicolumn{1}{c}{All\_zero\_2}      & \multicolumn{1}{c}{All\_one\_2}       & \multicolumn{1}{c|}{Random\_2}       & \multicolumn{1}{c}{All\_zero\_2} & \multicolumn{1}{c}{All\_one\_2} & \multicolumn{1}{c|}{Random\_2} & All\_zero\_2                          & All\_one\_2                           & Random\_2                             & All\_zero\_2                & All\_one\_2                  & Random\_2                   \\ \hline
\multicolumn{1}{c|}{All\_zero\_1} & {\color[HTML]{333333} 0.37}           & {\color[HTML]{FE0000} \textbf{16.34}} & {\color[HTML]{FE0000} \textbf{8.11}} & 1.72                             & 2.18                            & 1.88                           & {\color[HTML]{333333} 1.76}           & {\color[HTML]{FE0000} \textbf{19.74}} & {\color[HTML]{FE0000} \textbf{13.85}} & {\color[HTML]{000000} 0.65} & {\color[HTML]{000000} 0.54} & {\color[HTML]{000000} 0.18} \\
\multicolumn{1}{c|}{All\_one\_1}  & {\color[HTML]{FE0000} \textbf{17.09}} & {\color[HTML]{333333} 0.76}           & {\color[HTML]{FE0000} \textbf{8.64}} & 3.50                             & 3.97                            & 3.66                           & {\color[HTML]{FE0000} \textbf{17.67}} & {\color[HTML]{333333} 0.85}           & {\color[HTML]{FE0000} \textbf{5.04}}  & {\color[HTML]{000000} 1.39} & {\color[HTML]{000000} 1.30}  & {\color[HTML]{000000} 0.55} \\
Random\_1                         & {\color[HTML]{FE0000} \textbf{8.18}}  & {\color[HTML]{FE0000} \textbf{7.60}}  & {\color[HTML]{333333} 0.27}          & 0.48                             & 0.95                            & 0.65                           & {\color[HTML]{FE0000} \textbf{11.66}} & {\color[HTML]{FE0000} \textbf{7.27}}  & {\color[HTML]{333333} 1.12}           & {\color[HTML]{000000} 0.08} & {\color[HTML]{000000} 0.05}  & {\color[HTML]{000000} 0.76}
\end{tabular}%
}
\vspace*{-2mm}
\end{table*}

\subsection{User-Space Attack} \label{ssec:user-space}
We studied the feasibility of a user-space attack following the threat model assumed in Attack Scenario 2 (see Section~\ref{sssec:threat_model}). 
A user-space attacker also follows the attack methodology as presented in section~\ref{sssec:attack_method}. However, unlike a privileged attacker,  
a user-space attacker does not have the privilege to configure the reactive limits and hence, has to utilize the existing reactive limit settings.
This results in two implications: first, for most of the time, the attacker can only leverage the PL1 limit (or PPT limit on AMD system) to trigger frequency throttling as the default PL1 limit is typically set lower than the other reactive limits. 
However, PL1 limit usually has a much larger default $\tau$ window (in the order of tens of seconds) compared to that of PL2 limit, and hence, trace collection time due to PL1 limit-induced throttling will be significantly longer as well (see section \ref{ssec:online}).
Second, it is possible that the power consumption of a victim workload does not exceed even the PL1 limit, thus failing to trigger any throttling activity, requiring the adversary to compensate.

In order to overcome the limitations of the second implication, the user-space attacker can execute a {\em suitable} stressor code in parallel with the victim workload to boost the cumulative power consumption 
$\overline{P}$ such that it exceeds a reactive limit of the system. 
According to Algorithm \ref{algo:new_p_state} (line 8), a significant boost in $\overline{P}$ will lead to a reduction of the available power budget $\Delta$ below a certain threshold, which in turn will trigger frequency throttling activity.
Stressor code selection requires the following careful considerations:
first, power consumption of the stressor code should be high enough to aid the workload to hit a system reactive limit. Second, the variance in power consumption of the stressor code should be as low as possible, so that it does not introduce significant additional noise in the collected side-channel traces.
Third, the stressor code should have minimal resource contention (e.g., contention for ports or execution units) with the victim workload, to further reduce noise.
Additionally, the execution of the stressor code should not 
cause the system temperature to exceed the thermal limit, since in that case the collected side-channel traces will be affected by noise introduced due to thermal throttling.  
Therefore, as evident from the above requirements, the selection procedure of a suitable stressor code for a given victim workload and system under test could be a challenging task for the user-space attacker.

We performed experimental evaluations for the user-space attack scenario on an Intel i7-10700K system using its available 8 physical cores/16 logical cores (with SMT enabled).
At first, we profiled the power consumption $P_{victim}$ of victim workload (AES-NI based AES implementation) by simultaneously executing multiple instances of it on 8 different physical cores and observed that $P_{victim}$ is about 83W.
Next, we considered the following four test cases by assuming the default PL1 limit at different levels (but with the same $\tau$ window of about 56s):
\begin{enumerate}[leftmargin=*]
\item PL1 limit set to 80W. Only victim workload is executed.
\item PL1 limit set to 110W. Only victim workload is executed.
\item PL1 limit set to 110W. Stressor and victim workloads executed.
\item PL1 limit set to 140W. Stressor and victim workloads executed.
\end{enumerate}
\noindent For all the above cases, we performed TVLA tests,
by collecting 1,000 timing side-channel traces (with $T_{\delta}$ = 1s) corresponding to the encryption of each of the plaintext sets as considered previously.
Note that in these tests, we set $T_{\delta}$ to a much larger value (compared to 45ms as used in privileged attacker scenario) due to the high default value of PL1 $\tau$ window.
Additionally, for cases (3) and (4), a stressor code consisted of repeated execution of \texttt{MOVB} instructions across 6 different threads was running in parallel with the victim.

The results of the test cases are reported in Table~\ref{tab:PL1-TVLA} from left to right. 
For test case (1), when PL1=80W and is lower than $P_{victim}$, we observe high t-score values between different pairs of plaintext sets, signifying significant throttling side-channel leakage from the victim workload execution even without stressor code.
For test case (2), when PL1=110W and without any stressor code, we do not observe any leakage after analyzing the collected timing traces corresponding to the execution of victim workload. This is an expected outcome, since in this case the limit is higher and $P_{victim}$ fails to trigger PL1-induced frequency throttling activity. This was the reason for introducing stressor code. 
For test case (3), when the stressor code was executed in parallel to the victim workload, the cumulative system power consumption $\overline{P}$ hit the PL1 limit, leading to throttling side-channel leakage, as evident from the corresponding high t-score values in the third column. This shows that the stressor code is effective in inducing throttling side-channel leakage.
For test case (4), when PL1=140W, we observed that the simultaneous execution of the victim and stressor workloads caused system temperature to rise to 100$^\circ$C, thus triggering thermal throttling before triggering PL1-based throttling. 
Thermal throttling degraded the SNR of collected timing traces, thus leading to low t-score values as reported in the last column of Table~\ref{tab:PL1-TVLA}.

In this paper, for the user-space attack scenario, we limited our experimental evaluations to TVLA tests, as it was infeasible to perform a CPA attack against an AES-NI based AES implementation, because of its very high MTD requirements for collecting millions of telemetry side-channel traces. Importantly, in case of side-channel analysis of cryptosystems which require low number of traces, a user-space attack exploiting frequency throttling activity can be realized with a reasonable MTD.
For example, in~\cite{wan2022hertzbleed}, the authors demonstrate how a Ring 3 attacker utilizes the throttling side-channel information to extract keys from a post-quantum key encapsulation scheme within a few days, exploiting a corner case in the algorithm implementation.
\section{Mitigation} \label{sec:mitigation}
In this section, we discuss different countermeasures to safeguard a victim workload from being susceptible to frequency throttling side-channel attacks.
Before going to the details of the mitigation strategies, we first summarize the conditions that must be satisfied to mount such an attack.
\begin{itemize}[leftmargin=*]
\item {\bf Condition 1 (Secret Dependency):} The victim code processes a secret asset that is vulnerable to a power side-channel attack. This requires (i) the victim software implementation to be vulnerable to {\em traditional} physical side-channel attacks and (ii) the underlying hardware system to exhibit variation in power consumption profiles for processing different data.
\item {\bf Condition 2 (Controller Actuation):} One of the reactive limits of the system is being hit during victim code execution. This will cause the power management architecture to trigger frequency throttling activity based on the available power budget (see Algorithm~\ref{algo:new_p_state} for details).
\item {\bf Condition 3 (Observability):} The attacker can monitor the execution time (wall clock time) of the victim code with sufficiently high resolution, or else an equivalent quantity.
\end{itemize}

\noindent In order to thwart side-channel information leakage due to frequency throttling activity, the designer should consider targeting the above-mentioned necessary conditions. Next, we present different potential countermeasure options along with their respective advantages and disadvantages.

\begin{table*}[]
\centering
\caption{Summary of mitigations against the frequency throttling side-channel}
\label{tab:mitigation}
\vspace*{-3mm}
\resizebox{\textwidth}{!}{%
\begin{tabular}{l|l|l|l|l|l}
\# & Mitigation Target & Mitigation Description                                                   & Applicable Layer(s) & Mitigation Effectiveness         & Perf./Func. Impact            \\ \hline
1  & Secret Dependency        & Existing traditional power side-channel mitigations (e.g., masking, key refresh) & User App/System SW/HW   & Vary                             & Vary                          \\ \cline{2-6} 
2  & Controller Actuation        & Keep the system at lowest frequency or disable reactive limits                  & System SW/HW          & {\color[HTML]{32CB00} Fully}     & {\color[HTML]{FE0000} High}   \\
3  &                     & Reduce sensitivity of throttling control algorithm                              & HW                 & {\color[HTML]{F8A102} Partially} & {\color[HTML]{F8A102} Medium} \\
4  &                     & Add randomness to the reactive limit                                            & System SW/HW          & {\color[HTML]{F8A102} Partially} & {\color[HTML]{F8A102} Medium} \\
5  &                     & Avoid exposing reactive limit configuration interfaces to untrusted entities & System SW             & {\color[HTML]{F8A102} Partially} & {\color[HTML]{32CB00} Low}    \\
6  &                     & Use modelled power instead of actual power in throttling control algorithm      & HW                 & {\color[HTML]{32CB00} Fully}     & {\color[HTML]{FE0000} High}   \\
7  &                     & Add noise to power input of throttling control algorithm                        & HW                 & {\color[HTML]{F8A102} Partially} & {\color[HTML]{F8A102} Medium} \\ \cline{2-6} 
8  & Observability         & Utilize inherent noise or inject artificial noise to cryptographic operations   & User App/System SW      & {\color[HTML]{F8A102} Partially} & {\color[HTML]{F8A102} Medium}
\end{tabular}%
}
\vspace*{-2mm}
\end{table*}

\subsection{Analysis of Secret Dependency}
\subsubsection{Necessary conditions of Secret Dependency}
Since power side-channel is the fundamental root cause of frequency throttling side-channel, the necessary conditions for the physical power side-channel attack also need to be satisfied for throttling side-channel (except for the physical access capability to measure power). First, the victim application needs to process a secret asset (e.g., cryptographic key) with a confidentiality requirement. Second, power consumption of the underlying hardware processing the secret needs to correlate with the asset. Third, the implementation of the victim application is susceptible to power side-channel attack. For example, the victim application provides the capability for the adversary to repeatedly initiate cryptographic operations with the same sensitive key to collect enough data. Also, for block ciphers, the adversary should have the ability to read input/output or inter-round state of the block cipher primitives. Please note that the input/output is not necessarily the plaintext or ciphertext. One example is the counter (CTR) mode of operation for block ciphers, where the input to the block cipher is the concatenation of the nonce and the counter instead of a plaintext.

\subsubsection{Mitigations}
Most of the existing countermeasures against traditional power side-channel will be as effective against a frequency throttling side-channel. For example, software-based masking \cite{Prouff13EUROCRYPT} that splits a secret asset into multiple random shares will randomize the power consumption of the hardware, and will be useful against a frequency throttling side-channel.
There are several noteworthy exceptions. For example, shuffling-based countermeasures that randomize instruction execution order, while being effective in making trace alignment and identification of points of interest harder for physical power side-channel attacks, are less effective in mitigating a frequency throttling side-channel. This is because reordering instructions at the cycle granularity is less likely to impact average power consumption during the time window $\tau$ in the order or milliseconds or longer. Also, since an adversary does not need to physically access the hardware, any protection that physically isolates the system will not be sufficient to prevent a frequency throttling side-channel attack.

For cryptographic applications based on existing cryptographic libraries, an example of a generic countermeasure against power side-channel is \emph{key refresh}. One of the necessary conditions for power side-channel is \emph{amplification}, or the ability to repeatedly kick off cryptographic operations with the same sensitive key to collect a sufficient amount of traces. If the secret key is refreshed before enough traces can be collected, it will be harder for the attacker to fully deduce the secret. One important design factor is how frequency the key should be refreshed, which could be based on timing (e.g., refresh per several hours) or data volume (e.g., the volume of data being encrypted with the same key). If the designer is uncertain of the threshold to use, the lowest threshold that meets performance and design requirements should be selected. Naturally, the practicality of key refresh depends on the specific cryptographic use case (e.g., key refresh is typically not applicable to disk encryption).

From a hardware perspective, secret dependency is satisfied for almost all modern CPUs since power consumption difference due to circuit switching behavior is an inherent property of CMOS circuits. Making the entire SoC power-constant would of course address this condition, but is difficult to achieve, if not impossible. A more feasible option is making specific security-sensitive hardware components (e.g., hardware cryptography accelerator) power-constant.

\subsection{Analysis of Controller Actuation}
This condition allows conversion from power differences to timing differences during an attack. As described in Algorithm \ref{algo:new_p_state}, the new frequency limit after throttling $f_{max}$ is a function of the power budget, which is the difference between the power limit $PL$ and the measured average power consumption $\overline{P}$. 
A mitigation may target one of the components in this conversion process: the control algorithm, $PL$, or $\overline{P}$. 
\subsubsection{Mitigations targeting the control algorithm}
Since the purpose of reactive limits is to restrict a system from consuming power or current beyond the limit while maximizing system performance, a control algorithm is typically designed to select the highest possible frequency limit that satisfies the reactive limits. A straightforward mitigation option to change the control algorithm is to only allow the system to run at the lowest frequency when reactive limits are hit. While it prevents data-dependent frequency change, system performance is severely impacted. Another option is to fully disable reactive limit based throttling, which is usually not acceptable, since it is a critical power management feature and is required for safety. An option to trade-off between security and functionality is to reduce sensitivity of the control algorithm so that the switching of frequency limit would be less correlated with the input.
\subsubsection{Mitigations targeting reactive limits}
Similarly, a firmware or system software may take a straightforward approach to either configure the limit to a value too high to hit, or keep it very low so that the system always runs at the lowest frequency. 
However, these changes have severe negative impact on performance or functionality. One alternative solution is to randomly "fuzz'' the reactive limit. For example, instead of configuring a static reactive limit to $PL$, firmware or system software may define a range $[PL_{Low}, PL_{High}]$, and randomly select a value in the range, dynamically and routinely configuring the reactive limit. By doing so, randomness will be introduced in the power budget, as well as CPU frequency.

As discussed, interfaces (e.g., MSRs) to configure reactive limits, if accessible, could be utilized by an adversary to reduce the limits and trigger the throttling side-channel attack. A cloud service provider (CSP) or system software could prevent these interfaces from being exposed to untrusted guest VMs or ring-3 software, and be aware of the risk if the interfaces have to be exposed.
\subsubsection{Mitigations targeting average power}
The processor may decouple the calculated average power consumption from the actual power consumption. One approach is to utilize modelled power consumption instead of the actual power reading in the algorithm. If the model is selected to exclude information of instruction operands, then the average power will be independent of any secret data consumed by the victim application. Another approach is for the processor to "fuzz'' the average power consumption by adding noise to the value before the control algorithm uses it to compute the power budget. This is equivalent to the idea of fuzzing the reactive limit, since power budget is the difference between reactive limits and the average power. Please note that although fuzzing the power reading will not directly change power consumption, it will alter $f_{max}$ and indirectly impact power consumption and performance.

\subsection{Analysis of Observability}

One of the common countermeasures against side-channel attacks is to jam the channel with noise to prevent the attacker from deducing the secret. As the side-channel in this attack is frequency and timing information, noise can be injected into the frequency transition or timing information. 
One method is to leverage inherent noise during cryptographic application calls. As the cryptographic library provider or cryptographic application provider, one may restrict the maximal size allowed of processed data per API invocation, so that more invocations of the API are needed to process the same amount of data, and larger intrinsic noise will be introduced. 
Besides that, a cryptography implementer may proactively inject random noise to cryptographic operations to increase timing variation. To implement this countermeasure, the developer may add dummy instructions that introduce sufficient power or latency variation. The dummy instructions should be independent of the secret data used in the cryptographic function. For example, timing variation can be introduced using a loop of instructions with random iterations. In addition to that, any power variation induced by the dummy instructions may also increase the entropy of the frequency transition. To ensure randomness is introduced for every frequency transition, it is recommended that some noise is injected during every time window $\tau$ of the reactive limits that the attacker would target. One possible way to trade-off security and performance impact is to combine this scheme with a key refresh countermeasure, to increase the time needed to perform a successful attack to a key lifetime that is acceptable.

\subsection{Summary of Mitigation Options}
A summary of the mitigation options is listed in Table \ref{tab:mitigation}, categorized based on the condition to address, the layer(s) to apply, the security effectiveness in mitigating the frequency throttling side-channel, and the performance or functional impact. As can be seen, options that fully resolve the security issue (e.g., \#2 and \#6) bring high performance or functional impact, while options that partially reduce the security risk have low to medium impact. Depending on the layer in which the mitigation is applied, different options might be selected. For example, the developer of a user-space cryptography implementation may consider options \#1 and \#8, which are the options available to Ring 3 software.

\section{Related Work}\label{sec:related}

Related work has demonstrated the vulnerability of software     accessible energy and power telemetry information to side-channel analysis attacks.
In~\cite{Yan2015ASO}, the authors highlighted the use of software-accessible battery data of an Android phone to extract sensitive information from multiple applications.
It has been shown in~\cite{Fusi:masterthesis} that energy meter readings can be used to infer control flow dependency as well as cache hit/miss patterns for a program.
\cite{mantel2018secure} demonstrated the utilization of energy meter readings to distinguish keys of different Hamming weights for an RSA implementation. 

In addition to the above, several recent works have targeted a processor's energy consumption information (as exposed by RAPL interfaces in both Intel and AMD processors) to perform side-channel analysis attacks.
In~\cite{Lipp2020Platypus}, the authors demonstrate software-based power side-channel attacks called {\em Platypus} to extract a key from a secure enclave, to break kernel address space layout randomization (KASLR), and to establish a timing-independent covert channel. In~\cite{liu2021methodology}, the authors present a methodology to perform side-channel risk assessment of different software-accessible telemetries including RAPL energy, CPU frequency, voltage, and temperature data. 

Concurrent to our work, Wang \emph{et al.} \cite{wan2022hertzbleed} also discovered the frequency throttling side-channel and named the attack \emph{Hertzbleed}. Wang \emph{et al.} highlight the concept of conversion of power side-channel information to timing side-channel information and exploit it to recover a key from the SIKE post-quantum key encapsulation algorithm. Compared to Hertzbleed, this paper introduces and comprehensively explains the underlying mechanism and reasoning behind the frequency throttling side-channel leakage. We demonstrate key recovery by performing correlation power analysis on the widely used AES-NI based AES implementation and further elaborate on the privileged software attack scenario (sections~\ref{ssec:eval} and~\ref{ssec:intel_sgx}), showing how reactive limit configuration helps coerce information leakage. We discuss the unprivileged attack scenario, which the Hertzbleed paper focuses on, showing that such an attack is comparatively harder to achieve (section \ref{ssec:user-space}). In addition, this paper provides a thorough discussion of mitigation options to thwart such frequency throttling side-channel attacks. Wang \emph{et al.} propose to disable Turbo Boost or to disable SpeedStep and HWP from the BIOS, as a workload independent mitigation, with the presumed intent of minimizing DVFS transitions. While this has the effect of indirectly reducing power consumption, such a mitigation does not affect the enforcement of reactive limits nor the underlying behavior of Algorithm \ref{algo:new_p_state}. Thus, if a reactive limit is hit (e.g., as in the privileged attack scenario, by an adversary sufficiently lowering the reactive limit), then information is still expected to leak. 

\vspace*{-0.7mm}
\section{Conclusion} \label{sec:conclusion}
In this paper, we present a novel frequency throttling side-channel analysis attack. The root cause of such a side-channel is a power side-channel, which is converted to a timing side-channel by the power management architecture when reactive limits are triggered. 
We demonstrate the threat posed by frequency throttling side-channel attacks by showing that the cryptographic key can be successfully extracted from a constant-cycle implementation of AES by measuring the executing time of cryptographic operations and apply correlation power analysis.
Finally, we present a set of options to thwart such throttling side-channel analysis attacks, with analysis of pros and cons.
These mitigation options provide insights into the necessary conditions for throttling side-channel information leakage and how to develop effective countermeasures.

\bibliographystyle{ACM-Reference-Format}
\bibliography{references}


\begin{thebibliography}{00}


\ifx \showCODEN    \undefined \def \showCODEN     #1{\unskip}     \fi
\ifx \showDOI      \undefined \def \showDOI       #1{#1}\fi
\ifx \showISBNx    \undefined \def \showISBNx     #1{\unskip}     \fi
\ifx \showISBNxiii \undefined \def \showISBNxiii  #1{\unskip}     \fi
\ifx \showISSN     \undefined \def \showISSN      #1{\unskip}     \fi
\ifx \showLCCN     \undefined \def \showLCCN      #1{\unskip}     \fi
\ifx \shownote     \undefined \def \shownote      #1{#1}          \fi
\ifx \showarticletitle \undefined \def \showarticletitle #1{#1}   \fi
\ifx \showURL      \undefined \def \showURL       {\relax}        \fi
\providecommand\bibfield[2]{#2}
\providecommand\bibinfo[2]{#2}
\providecommand\natexlab[1]{#1}
\providecommand\showeprint[2][]{arXiv:#2}

\bibitem[\protect\citeauthoryear{AMD}{AMD}{2014}]%
        {Ryzen_monitor}
\bibfield{author}{\bibinfo{person}{AMD}.} \bibinfo{year}{2014}\natexlab{}.
\newblock \bibinfo{title}{Ryzen\_Monitor}.
\newblock
  \bibinfo{howpublished}{\url{https://github.com/hattedsquirrel/ryzen\_monitor}}.
    (\bibinfo{year}{2014}).
\newblock
\newblock
\shownote{Accessed: 2022-05-01.}


\bibitem[\protect\citeauthoryear{AMD}{AMD}{2018a}]%
        {kernel-developer-guide-AMD-15H}
\bibfield{author}{\bibinfo{person}{AMD}.} \bibinfo{year}{2018}\natexlab{a}.
\newblock \bibinfo{title}{BIOS and Kernel Developer’s Guide (BKDG) for AMD
  Family 15h Models 70h-7Fh Processors}.
\newblock
  \bibinfo{howpublished}{\url{https://www.amd.com/system/files/TechDocs/55072\_AMD\_Family\_15h\_Models\_70h-7Fh\_BKDG.pdf}}.
    (\bibinfo{year}{2018}).
\newblock
\newblock
\shownote{Accessed: 2022-05-01.}


\bibitem[\protect\citeauthoryear{AMD}{AMD}{2018b}]%
        {Open-Source-Register-Reference-AMD-17H}
\bibfield{author}{\bibinfo{person}{AMD}.} \bibinfo{year}{2018}\natexlab{b}.
\newblock \bibinfo{title}{Open-Source Register Reference For AMD Family 17h
  Processors Models 00h-2Fh}.
\newblock
  \bibinfo{howpublished}{\url{https://developer.amd.com/wp-content/resources/56255_3_03.PDF}}.
    (\bibinfo{year}{2018}).
\newblock
\newblock
\shownote{Accessed: 2022-05-01.}


\bibitem[\protect\citeauthoryear{AMD}{AMD}{2022a}]%
        {AMD-RAPL-Patch}
\bibfield{author}{\bibinfo{person}{AMD}.} \bibinfo{year}{2022}\natexlab{a}.
\newblock \bibinfo{title}{AMD CVE-2020-12912}.
\newblock
  \bibinfo{howpublished}{\url{https://nvd.nist.gov/vuln/detail/CVE-2020-12912}}.
    (\bibinfo{year}{2022}).
\newblock
\newblock
\shownote{Accessed: 2022-04-12.}


\bibitem[\protect\citeauthoryear{AMD}{AMD}{2022b}]%
        {AMD-PB2}
\bibfield{author}{\bibinfo{person}{AMD}.} \bibinfo{year}{2022}\natexlab{b}.
\newblock \bibinfo{title}{AMD Ryzen Technology: Precision Boost 2 Performance
  Enhancement}.
\newblock
  \bibinfo{howpublished}{\url{https://www.amd.com/en/support/kb/faq/cpu-pb2}}.
   (\bibinfo{year}{2022}).
\newblock
\newblock
\shownote{Accessed: 2022-03-12.}


\bibitem[\protect\citeauthoryear{AMD}{AMD}{2022c}]%
        {AMD-user-guide}
\bibfield{author}{\bibinfo{person}{AMD}.} \bibinfo{year}{2022}\natexlab{c}.
\newblock \bibinfo{title}{{AMD} u{Prof} {User} {Guide}}.
\newblock
  \bibinfo{howpublished}{\url{https://developer.amd.com/wordpress/media/2013/12/User_Guide.pdf}}.
    (\bibinfo{year}{2022}).
\newblock
\newblock
\shownote{Accessed: 2022-4-21.}


\bibitem[\protect\citeauthoryear{AMD}{AMD}{2022d}]%
        {amd-ryzen-master}
\bibfield{author}{\bibinfo{person}{AMD}.} \bibinfo{year}{2022}\natexlab{d}.
\newblock \bibinfo{title}{Ryzen Master 2.9 - Reference Guide}.
\newblock
  \bibinfo{howpublished}{\url{https://www.amd.com/system/files/documents/ryzen-master-quick-reference-guide.pdf}}.
    (\bibinfo{year}{2022}).
\newblock
\newblock
\shownote{Accessed: 2022-05-01.}


\bibitem[\protect\citeauthoryear{ARM}{ARM}{2022}]%
        {ARM-PM}
\bibfield{author}{\bibinfo{person}{ARM}.} \bibinfo{year}{2022}\natexlab{}.
\newblock \bibinfo{title}{ARMv8-A Power Management}.
\newblock
  \bibinfo{howpublished}{\url{https://developer.arm.com/documentation/100960/0100/ARMv8-A-Power-management?lang=en}}.
    (\bibinfo{year}{2022}).
\newblock
\newblock
\shownote{Accessed: 2022-04-10.}


\bibitem[\protect\citeauthoryear{{\AA}str{\"o}m and Murray}{{\AA}str{\"o}m and
  Murray}{2010}]%
        {aastrom2010feedback}
\bibfield{author}{\bibinfo{person}{Karl~Johan {\AA}str{\"o}m} {and}
  \bibinfo{person}{Richard~M Murray}.} \bibinfo{year}{2010}\natexlab{}.
\newblock \showarticletitle{Feedback systems}.
\newblock In \bibinfo{booktitle}{{\em Feedback Systems}}.
  \bibinfo{publisher}{Princeton university press}.
\newblock


\bibitem[\protect\citeauthoryear{Ayers, Nagendra, August, Cho, Kanev,
  Kozyrakis, Krishnamurthy, Litz, Moseley, and Ranganathan}{Ayers
  et~al\mbox{.}}{2019}]%
        {AsmDB}
\bibfield{author}{\bibinfo{person}{Grant Ayers}, \bibinfo{person}{Nayana~Prasad
  Nagendra}, \bibinfo{person}{David~I. August}, \bibinfo{person}{Hyoun~Kyu
  Cho}, \bibinfo{person}{Svilen Kanev}, \bibinfo{person}{Christos Kozyrakis},
  \bibinfo{person}{Trivikram Krishnamurthy}, \bibinfo{person}{Heiner Litz},
  \bibinfo{person}{Tipp Moseley}, {and} \bibinfo{person}{Parthasarathy
  Ranganathan}.} \bibinfo{year}{2019}\natexlab{}.
\newblock \showarticletitle{AsmDB: Understanding and Mitigating Front-End
  Stalls in Warehouse-Scale Computers}. In \bibinfo{booktitle}{{\em Proceedings
  of the 46th International Symposium on Computer Architecture}} {\em
  (\bibinfo{series}{ISCA '19})}. \bibinfo{address}{New York, NY, USA},
  \bibinfo{pages}{462–473}.
\newblock
\showDOI{%
\url{https://doi.org/10.1145/3307650.3322234}}


\bibitem[\protect\citeauthoryear{Bircher and John}{Bircher and John}{2008}]%
        {bircher2008analysis}
\bibfield{author}{\bibinfo{person}{W.~Lloyd Bircher} {and}
  \bibinfo{person}{Lizy~K. John}.} \bibinfo{year}{2008}\natexlab{}.
\newblock \showarticletitle{Analysis of Dynamic Power Management on Multi-Core
  Processors}. In \bibinfo{booktitle}{{\em Proceedings of the 22nd Annual
  International Conference on Supercomputing}} {\em (\bibinfo{series}{ICS
  '08})}. \bibinfo{address}{New York, NY, USA}, \bibinfo{pages}{327–338}.
\newblock
\showDOI{%
\url{https://doi.org/10.1145/1375527.1375575}}


\bibitem[\protect\citeauthoryear{Colmant, Felber, Rouvoy, and
  Seinturier}{Colmant et~al\mbox{.}}{2017}]%
        {Colmant2017WattsKitSP}
\bibfield{author}{\bibinfo{person}{Maxime Colmant}, \bibinfo{person}{Pascal
  Felber}, \bibinfo{person}{Romain Rouvoy}, {and} \bibinfo{person}{Lionel
  Seinturier}.} \bibinfo{year}{2017}\natexlab{}.
\newblock \showarticletitle{WattsKit: Software-Defined Power Monitoring of
  Distributed Systems}.
\newblock \bibinfo{journal}{{\em 2017 17th IEEE/ACM International Symposium on
  Cluster, Cloud and Grid Computing (CCGRID)\/}} (\bibinfo{year}{2017}),
  \bibinfo{pages}{514--523}.
\newblock
\showDOI{%
\url{https://doi.org/10.1109/CCGRID.2017.27}}


\bibitem[\protect\citeauthoryear{Corporation and B}{Corporation and B}{2000}]%
        {acpi00specification}
\bibfield{author}{\bibinfo{person}{Compaq~Computer Corporation} {and}
  \bibinfo{person}{Revision B}.} \bibinfo{year}{2000}\natexlab{}.
\newblock \bibinfo{title}{Advanced Configuration and Power Interface
  Specification}.
\newblock   (\bibinfo{year}{2000}).
\newblock
\showURL{%
\url{http://www.acpi.info/}}


\bibitem[\protect\citeauthoryear{Costan and Devadas}{Costan and
  Devadas}{2016}]%
        {costan2016intel}
\bibfield{author}{\bibinfo{person}{Victor Costan} {and}
  \bibinfo{person}{Srinivas Devadas}.} \bibinfo{year}{2016}\natexlab{}.
\newblock \bibinfo{title}{Intel SGX Explained}.
\newblock \bibinfo{howpublished}{Cryptology ePrint Archive, Paper 2016/086}.
  (\bibinfo{year}{2016}).
\newblock
\showURL{%
\url{https://eprint.iacr.org/2016/086}}


\bibitem[\protect\citeauthoryear{Fieni, Rouvoy, and Seinturier}{Fieni
  et~al\mbox{.}}{2020}]%
        {Fieni2020SmartWattsSS}
\bibfield{author}{\bibinfo{person}{Guillaume Fieni}, \bibinfo{person}{Romain
  Rouvoy}, {and} \bibinfo{person}{Lionel Seinturier}.}
  \bibinfo{year}{2020}\natexlab{}.
\newblock \showarticletitle{SmartWatts: Self-Calibrating Software-Defined Power
  Meter for Containers}.
\newblock \bibinfo{journal}{{\em 2020 20th IEEE/ACM International Symposium on
  Cluster, Cloud and Internet Computing (CCGRID)\/}} (\bibinfo{year}{2020}),
  \bibinfo{pages}{479--488}.
\newblock
\showDOI{%
\url{https://doi.org/10.1109/CCGrid49817.2020.00-45}}


\bibitem[\protect\citeauthoryear{Fusi}{Fusi}{2016}]%
        {Fusi:masterthesis}
\bibfield{author}{\bibinfo{person}{Matteo~Maria Fusi}.}
  \bibinfo{year}{2016}\natexlab{}.
\newblock {\em \bibinfo{title}{{Information-Leakage Analysis based on Hardware
  Performance Counters}}}.
\newblock \bibinfo{thesistype}{Master's\ thesis}. \bibinfo{school}{The
  Polytechnic University of Milan}.
\newblock


\bibitem[\protect\citeauthoryear{Gilbert~Goodwill, Jaffe, Rohatgi,
  et~al\mbox{.}}{Gilbert~Goodwill et~al\mbox{.}}{2011}]%
        {gilbert2011testing}
\bibfield{author}{\bibinfo{person}{Benjamin~Jun Gilbert~Goodwill},
  \bibinfo{person}{Josh Jaffe}, \bibinfo{person}{Pankaj Rohatgi},
  {et~al\mbox{.}}} \bibinfo{year}{2011}\natexlab{}.
\newblock \showarticletitle{A testing methodology for side-channel resistance
  validation}. In \bibinfo{booktitle}{{\em NIST non-invasive attack testing
  workshop}}, Vol.~\bibinfo{volume}{7}.
\newblock


\bibitem[\protect\citeauthoryear{Goodwill, Jun, Jaffe, and Rohatgi}{Goodwill
  et~al\mbox{.}}{2011}]%
        {Goodwill11}
\bibfield{author}{\bibinfo{person}{Gilbert Goodwill}, \bibinfo{person}{Benjamin
  Jun}, \bibinfo{person}{Josh Jaffe}, {and} \bibinfo{person}{Pankaj Rohatgi}.}
  \bibinfo{year}{2011}\natexlab{}.
\newblock \bibinfo{title}{P.: A testing methodology for side‐channel
  resistance validation, NIAT}.
\newblock   (\bibinfo{year}{2011}).
\newblock


\bibitem[\protect\citeauthoryear{Gough, Steiner, and Saunders}{Gough
  et~al\mbox{.}}{2015}]%
        {gough2015cpu}
\bibfield{author}{\bibinfo{person}{Corey Gough}, \bibinfo{person}{Ian Steiner},
  {and} \bibinfo{person}{Winston Saunders}.} \bibinfo{year}{2015}\natexlab{}.
\newblock \bibinfo{booktitle}{{\em CPU power management}}.
\newblock \bibinfo{publisher}{Apress}, \bibinfo{address}{Berkeley, CA},
  \bibinfo{pages}{21--70}.
\newblock
\showDOI{%
\url{https://doi.org/10.1007/978-1-4302-6638-9_2}}


\bibitem[\protect\citeauthoryear{Gueron}{Gueron}{2010}]%
        {gueron2010intel}
\bibfield{author}{\bibinfo{person}{Shay Gueron}.}
  \bibinfo{year}{2010}\natexlab{}.
\newblock \bibinfo{title}{Intel{\textregistered} Advanced Encryption Standard
  (AES) New Instructions Set}.
\newblock   (\bibinfo{year}{2010}).
\newblock
\showURL{%
\url{https://www.intel.com/content/dam/doc/white-paper/advanced-encryption-standard-new-instructions-set-paper.pdf}}


\bibitem[\protect\citeauthoryear{Haj-Yahya, Mendelson, Ben-asher, and
  Chattopadhyay}{Haj-Yahya et~al\mbox{.}}{2018}]%
        {energyefficiencybook}
\bibfield{author}{\bibinfo{person}{Jawad Haj-Yahya}, \bibinfo{person}{Avi
  Mendelson}, \bibinfo{person}{Yosi Ben-asher}, {and} \bibinfo{person}{Anupam
  Chattopadhyay}.} \bibinfo{year}{2018}\natexlab{}.
\newblock \bibinfo{booktitle}{{\em Energy Efficient High Performance Processors
  Recent Approaches for Designing Green High Performance Computing}}.
\newblock
\showISBNx{978-981-10-8553-6}
\showDOI{%
\url{https://doi.org/10.1007/978-981-10-8554-3}}


\bibitem[\protect\citeauthoryear{Haj-Yahya, Orosa, Kim, Luna, Yaglikci, Alser,
  Puddu, and Mutlu}{Haj-Yahya et~al\mbox{.}}{2021}]%
        {IChannels}
\bibfield{author}{\bibinfo{person}{J. Haj-Yahya}, \bibinfo{person}{L. Orosa},
  \bibinfo{person}{J.~S. Kim}, \bibinfo{person}{J.~Gomez Luna},
  \bibinfo{person}{A. Yaglikci}, \bibinfo{person}{M. Alser},
  \bibinfo{person}{I. Puddu}, {and} \bibinfo{person}{O. Mutlu}.}
  \bibinfo{year}{2021}\natexlab{}.
\newblock \showarticletitle{IChannels: Exploiting Current Management Mechanisms
  to Create Covert Channels in Modern Processors}. In \bibinfo{booktitle}{{\em
  2021 ACM/IEEE 48th Annual International Symposium on Computer Architecture
  (ISCA)}}. \bibinfo{publisher}{IEEE Computer Society}, \bibinfo{address}{Los
  Alamitos, CA, USA}, \bibinfo{pages}{985--998}.
\newblock
\showDOI{%
\url{https://doi.org/10.1109/ISCA52012.2021.00081}}


\bibitem[\protect\citeauthoryear{Hochschild, Turner, Mogul, Govindaraju,
  Ranganathan, Culler, and Vahdat}{Hochschild et~al\mbox{.}}{2021}]%
        {MercurialCores}
\bibfield{author}{\bibinfo{person}{Peter~H. Hochschild}, \bibinfo{person}{Paul
  Turner}, \bibinfo{person}{Jeffrey~C. Mogul}, \bibinfo{person}{Rama
  Govindaraju}, \bibinfo{person}{Parthasarathy Ranganathan},
  \bibinfo{person}{David~E. Culler}, {and} \bibinfo{person}{Amin Vahdat}.}
  \bibinfo{year}{2021}\natexlab{}.
\newblock \showarticletitle{Cores That Don't Count}. In
  \bibinfo{booktitle}{{\em Proceedings of the Workshop on Hot Topics in
  Operating Systems}} {\em (\bibinfo{series}{HotOS '21})}.
  \bibinfo{address}{New York, NY, USA}, \bibinfo{pages}{9–16}.
\newblock
\showDOI{%
\url{https://doi.org/10.1145/3458336.3465297}}


\bibitem[\protect\citeauthoryear{Intel}{Intel}{2022a}]%
        {Intel-coding-guideline}
\bibfield{author}{\bibinfo{person}{Intel}.} \bibinfo{year}{2022}\natexlab{a}.
\newblock \bibinfo{title}{{G}uidelines for {M}itigating {T}iming {S}ide
  {C}hannels {A}gainst {C}ryptographic {I}mplementations}.
\newblock
  \bibinfo{howpublished}{\url{https://www.intel.com/content/www/us/en/developer/articles/technical/software-security-guidance/secure-coding/mitigate-timing-side-channel-crypto-implementation.html}}.
    (\bibinfo{year}{2022}).
\newblock
\newblock
\shownote{Accessed: 2022-05-02.}


\bibitem[\protect\citeauthoryear{Intel}{Intel}{2022b}]%
        {Intel-RAPL-Patch}
\bibfield{author}{\bibinfo{person}{Intel}.} \bibinfo{year}{2022}\natexlab{b}.
\newblock \bibinfo{title}{Intel CVE-2020-8694}.
\newblock
  \bibinfo{howpublished}{\url{https://www.intel.com/content/www/us/en/security-center/advisory/intel-sa-00389.html}}.
    (\bibinfo{year}{2022}).
\newblock
\newblock
\shownote{Accessed: 2022-04-12.}


\bibitem[\protect\citeauthoryear{Intel}{Intel}{2022c}]%
        {Intel-RAPL-Fuzz}
\bibfield{author}{\bibinfo{person}{Intel}.} \bibinfo{year}{2022}\natexlab{c}.
\newblock \bibinfo{title}{Intel Running Average Power Limit Energy Reporting}.
\newblock
  \bibinfo{howpublished}{\url{https://www.intel.com/content/www/us/en/developer/articles/technical/software-security-guidance/advisory-guidance/running-average-power-limit-energy-reporting.html}}.
    (\bibinfo{year}{2022}).
\newblock
\newblock
\shownote{Accessed: 2022-04-12.}


\bibitem[\protect\citeauthoryear{Intel}{Intel}{2022d}]%
        {SGX}
\bibfield{author}{\bibinfo{person}{Intel}.} \bibinfo{year}{2022}\natexlab{d}.
\newblock \bibinfo{title}{{I}ntel {SGX}}.
\newblock \bibinfo{howpublished}{\url{https://software.intel.com/en-us/sgx}}.
  (\bibinfo{year}{2022}).
\newblock
\newblock
\shownote{Accessed: 2022-05-02.}


\bibitem[\protect\citeauthoryear{Intel}{Intel}{2022e}]%
        {Intel-SDM}
\bibfield{author}{\bibinfo{person}{Intel}.} \bibinfo{year}{2022}\natexlab{e}.
\newblock \bibinfo{title}{{Intel® 64 and IA-32 Architectures Software
  Developer Manuals}}.
\newblock
  \bibinfo{howpublished}{\url{https://software.intel.com/content/www/us/en/develop/articles/intel-sdm.html}}.
    (\bibinfo{year}{2022}).
\newblock
\newblock
\shownote{Accessed: 2022-05-02.}


\bibitem[\protect\citeauthoryear{Intel}{Intel}{2022f}]%
        {ipp-crypto}
\bibfield{author}{\bibinfo{person}{Intel}.} \bibinfo{year}{2022}\natexlab{f}.
\newblock \bibinfo{title}{Intel® Integrated Performance Primitives
  Cryptography}.
\newblock \bibinfo{howpublished}{\url{https://github.com/intel/ipp-crypto}}.
  (\bibinfo{year}{2022}).
\newblock
\newblock
\shownote{Accessed: 2022-04-07.}


\bibitem[\protect\citeauthoryear{Intel}{Intel}{2022g}]%
        {Intel-SpeedStep}
\bibfield{author}{\bibinfo{person}{Intel}.} \bibinfo{year}{2022}\natexlab{g}.
\newblock \bibinfo{title}{Overview of Enhanced Intel SpeedStep® Technology for
  Intel® Processors}.
\newblock
  \bibinfo{howpublished}{\url{https://www.intel.com/content/www/us/en/support/articles/000007073/processors.html}}.
    (\bibinfo{year}{2022}).
\newblock
\newblock
\shownote{Accessed: 2022-04-07.}


\bibitem[\protect\citeauthoryear{Intel}{Intel}{2022h}]%
        {Intel-Speed-Shift}
\bibfield{author}{\bibinfo{person}{Intel}.} \bibinfo{year}{2022}\natexlab{h}.
\newblock \bibinfo{title}{Overview of Intel® Speed Shift Technology}.
\newblock
  \bibinfo{howpublished}{\url{https://edc.intel.com/content/www/us/en/design/ipla/software-development-platforms/client/platforms/alder-lake-desktop/12th-generation-intel-core-processors-datasheet-volume-1-of-2/002/intel-speed-shift-technology/}}.
    (\bibinfo{year}{2022}).
\newblock
\newblock
\shownote{Accessed: 2022-04-07.}


\bibitem[\protect\citeauthoryear{irusanov}{irusanov}{2021}]%
        {ZenStates-Core}
\bibfield{author}{\bibinfo{person}{irusanov}.} \bibinfo{year}{2021}\natexlab{}.
\newblock \bibinfo{title}{ZenStates-Core}.
\newblock
  \bibinfo{howpublished}{\url{https://github.com/irusanov/ZenStates-Core}}.
  (\bibinfo{year}{2021}).
\newblock
\newblock
\shownote{Accessed: 2022-07-25.}


\bibitem[\protect\citeauthoryear{Kaplan, Powell, and Woller}{Kaplan
  et~al\mbox{.}}{2016}]%
        {kaplan2016amd}
\bibfield{author}{\bibinfo{person}{David Kaplan}, \bibinfo{person}{Jeremy
  Powell}, {and} \bibinfo{person}{Tom Woller}.}
  \bibinfo{year}{2016}\natexlab{}.
\newblock \bibinfo{title}{AMD memory encryption}.
\newblock   (\bibinfo{year}{2016}).
\newblock
\showURL{%
\url{https://developer.amd.com/wordpress/media/2013/12/AMD_Memory_Encryption_Whitepaper_v7-Public.pdf}}
\newblock
\shownote{Accessed: 2022-09-06.}


\bibitem[\protect\citeauthoryear{Kim, Gupta, Wei, and Brooks}{Kim
  et~al\mbox{.}}{2008}]%
        {kim2008system}
\bibfield{author}{\bibinfo{person}{Wonyoung Kim}, \bibinfo{person}{Meeta~S
  Gupta}, \bibinfo{person}{Gu-Yeon Wei}, {and} \bibinfo{person}{David Brooks}.}
  \bibinfo{year}{2008}\natexlab{}.
\newblock \showarticletitle{System level analysis of fast, per-core DVFS using
  on-chip switching regulators}. In \bibinfo{booktitle}{{\em 2008 IEEE 14th
  International Symposium on High Performance Computer Architecture}}.
  \bibinfo{pages}{123--134}.
\newblock
\showDOI{%
\url{https://doi.org/10.1109/HPCA.2008.4658633}}


\bibitem[\protect\citeauthoryear{Kocher, Jaffe, Jun, and Rohatgi}{Kocher
  et~al\mbox{.}}{2011}]%
        {kocher2011introduction}
\bibfield{author}{\bibinfo{person}{Paul Kocher}, \bibinfo{person}{Joshua
  Jaffe}, \bibinfo{person}{Benjamin Jun}, {and} \bibinfo{person}{Pankaj
  Rohatgi}.} \bibinfo{year}{2011}\natexlab{}.
\newblock \showarticletitle{Introduction to differential power analysis}.
\newblock \bibinfo{journal}{{\em Journal of Cryptographic Engineering\/}}
  \bibinfo{volume}{1}, \bibinfo{number}{1} (\bibinfo{year}{2011}),
  \bibinfo{pages}{5--27}.
\newblock


\bibitem[\protect\citeauthoryear{Kogler, Gruss, and Schwarz}{Kogler
  et~al\mbox{.}}{2022}]%
        {kogler2022minefield}
\bibfield{author}{\bibinfo{person}{Andreas Kogler}, \bibinfo{person}{Daniel
  Gruss}, {and} \bibinfo{person}{Michael Schwarz}.}
  \bibinfo{year}{2022}\natexlab{}.
\newblock \showarticletitle{Minefield: A Software-only Protection for {SGX}
  Enclaves against {DVFS} Attacks}. In \bibinfo{booktitle}{{\em 31st USENIX
  Security Symposium (USENIX Security 22)}}. \bibinfo{address}{Boston, MA},
  \bibinfo{pages}{4147--4164}.
\newblock


\bibitem[\protect\citeauthoryear{leogx9r}{leogx9r}{2022}]%
        {Ryzen_SMU}
\bibfield{author}{\bibinfo{person}{leogx9r}.} \bibinfo{year}{2022}\natexlab{}.
\newblock \bibinfo{title}{Ryzen {SMU}}.
\newblock \bibinfo{howpublished}{\url{https://gitlab.com/leogx9r/ryzen_smu}}.
  (\bibinfo{year}{2022}).
\newblock


\bibitem[\protect\citeauthoryear{Lipp, Kogler, Oswald, Schwarz, Easdon,
  Canella, and Gruss}{Lipp et~al\mbox{.}}{2021}]%
        {Lipp2020Platypus}
\bibfield{author}{\bibinfo{person}{Moritz Lipp}, \bibinfo{person}{Andreas
  Kogler}, \bibinfo{person}{David Oswald}, \bibinfo{person}{Michael Schwarz},
  \bibinfo{person}{Catherine Easdon}, \bibinfo{person}{Claudio Canella}, {and}
  \bibinfo{person}{Daniel Gruss}.} \bibinfo{year}{2021}\natexlab{}.
\newblock \showarticletitle{{PLATYPUS: Software-based Power Side-Channel
  Attacks on x86}}. In \bibinfo{booktitle}{{\em 2021 IEEE Symposium on Security
  and Privacy (SP)}}. \bibinfo{pages}{355--371}.
\newblock


\bibitem[\protect\citeauthoryear{Liu, Kar, Wang, Chawla, Roggel, Yuce, and
  Fung}{Liu et~al\mbox{.}}{2021}]%
        {liu2021methodology}
\bibfield{author}{\bibinfo{person}{Chen Liu}, \bibinfo{person}{Monodeep Kar},
  \bibinfo{person}{Xueyang Wang}, \bibinfo{person}{Nikhil Chawla},
  \bibinfo{person}{Neer Roggel}, \bibinfo{person}{Bilgiday Yuce}, {and}
  \bibinfo{person}{Jason~M Fung}.} \bibinfo{year}{2021}\natexlab{}.
\newblock \showarticletitle{Methodology of Assessing Information Leakage
  through Software-Accessible Telemetries}. In \bibinfo{booktitle}{{\em 2021
  IEEE International Symposium on Hardware Oriented Security and Trust
  (HOST)}}. IEEE, \bibinfo{pages}{259--269}.
\newblock


\bibitem[\protect\citeauthoryear{Mangard, Oswald, and Popp}{Mangard
  et~al\mbox{.}}{2008}]%
        {mangard2008power}
\bibfield{author}{\bibinfo{person}{Stefan Mangard}, \bibinfo{person}{Elisabeth
  Oswald}, {and} \bibinfo{person}{Thomas Popp}.}
  \bibinfo{year}{2008}\natexlab{}.
\newblock \bibinfo{booktitle}{{\em Power analysis attacks: Revealing the
  secrets of smart cards}}. Vol.~\bibinfo{volume}{31}.
\newblock \bibinfo{publisher}{Springer Science \& Business Media}.
\newblock


\bibitem[\protect\citeauthoryear{Mantel, Schickel, Weber, and Weber}{Mantel
  et~al\mbox{.}}{2018}]%
        {mantel2018secure}
\bibfield{author}{\bibinfo{person}{Heiko Mantel}, \bibinfo{person}{Johannes
  Schickel}, \bibinfo{person}{Alexandra Weber}, {and}
  \bibinfo{person}{Friedrich Weber}.} \bibinfo{year}{2018}\natexlab{}.
\newblock \showarticletitle{How Secure Is Green IT? The Case of Software-Based
  Energy Side Channels}. In \bibinfo{booktitle}{{\em European Symposium on
  Research in Computer Security (ESORICS 2018)}}. \bibinfo{pages}{218--239}.
\newblock


\bibitem[\protect\citeauthoryear{Mestiri, Benhadjyoussef, Machhout, and
  Tourki}{Mestiri et~al\mbox{.}}{2012}]%
        {HW-HD-power-model}
\bibfield{author}{\bibinfo{person}{Hassen Mestiri}, \bibinfo{person}{Noura
  Benhadjyoussef}, \bibinfo{person}{Mohsen Machhout}, {and}
  \bibinfo{person}{Rached Tourki}.} \bibinfo{year}{2012}\natexlab{}.
\newblock \showarticletitle{A Comparative Study of Power Consumption Models for
  CPA Attack}.
\newblock \bibinfo{journal}{{\em International Journal of Computer Network and
  Information Security\/}}  \bibinfo{volume}{5} (\bibinfo{date}{03}
  \bibinfo{year}{2012}), \bibinfo{pages}{25--31}.
\newblock
\showDOI{%
\url{https://doi.org/10.5815/ijcnis.2013.03.03}}


\bibitem[\protect\citeauthoryear{Morbitzer, Proskurin, Radev, Dorfhuber, and
  Salas}{Morbitzer et~al\mbox{.}}{2021}]%
        {SEVerity}
\bibfield{author}{\bibinfo{person}{Mathias Morbitzer}, \bibinfo{person}{Sergej
  Proskurin}, \bibinfo{person}{Martin Radev}, \bibinfo{person}{Marko
  Dorfhuber}, {and} \bibinfo{person}{Erick~Quintanar Salas}.}
  \bibinfo{year}{2021}\natexlab{}.
\newblock \showarticletitle{SEVerity: Code Injection Attacks against Encrypted
  Virtual Machines}. In \bibinfo{booktitle}{{\em 2021 IEEE Security and Privacy
  Workshops (SPW)}}. \bibinfo{pages}{444--455}.
\newblock
\showDOI{%
\url{https://doi.org/10.1109/SPW53761.2021.00063}}


\bibitem[\protect\citeauthoryear{Muralidhar, Borovica-Gajic, and
  Buyya}{Muralidhar et~al\mbox{.}}{2022}]%
        {EnergyEfficientComputing}
\bibfield{author}{\bibinfo{person}{Rajeev Muralidhar}, \bibinfo{person}{Renata
  Borovica-Gajic}, {and} \bibinfo{person}{Rajkumar Buyya}.}
  \bibinfo{year}{2022}\natexlab{}.
\newblock \showarticletitle{Energy Efficient Computing Systems: Architectures,
  Abstractions and Modeling to Techniques and Standards}.
\newblock \bibinfo{journal}{{\it Comput. Surveys}} (\bibinfo{date}{Jan.}
  \bibinfo{year}{2022}).
\newblock
\showDOI{%
\url{https://doi.org/10.1145/3511094}}


\bibitem[\protect\citeauthoryear{Noureddine, Rouvoy, and Seinturier}{Noureddine
  et~al\mbox{.}}{2015}]%
        {MonitoringEnergyHotspots}
\bibfield{author}{\bibinfo{person}{Adel Noureddine}, \bibinfo{person}{Romain
  Rouvoy}, {and} \bibinfo{person}{Lionel Seinturier}.}
  \bibinfo{year}{2015}\natexlab{}.
\newblock \showarticletitle{Monitoring Energy Hotspots in Software}.
\newblock \bibinfo{journal}{{\em Automated Software Engg.\/}}
  \bibinfo{volume}{22}, \bibinfo{number}{3} (\bibinfo{date}{sep}
  \bibinfo{year}{2015}), \bibinfo{pages}{291–332}.
\newblock
\showISSN{0928-8910}
\showDOI{%
\url{https://doi.org/10.1007/s10515-014-0171-1}}


\bibitem[\protect\citeauthoryear{of~Standards and Technology}{of~Standards and
  Technology}{2001}]%
        {FIPS197}
\bibfield{author}{\bibinfo{person}{National~Institute of Standards} {and}
  \bibinfo{person}{Technology}.} \bibinfo{year}{2001}\natexlab{}.
\newblock \bibinfo{booktitle}{{\em ADVANCED ENCRYPTION STANDARD (AES)}}.
\newblock \bibinfo{type}{{T}echnical {R}eport}. \bibinfo{institution}{U.S.
  Department of Commerce}, \bibinfo{address}{Washington, D.C.}
\newblock


\bibitem[\protect\citeauthoryear{Prouff and Rivain}{Prouff and Rivain}{2013}]%
        {Prouff13EUROCRYPT}
\bibfield{author}{\bibinfo{person}{Emmanuel Prouff} {and}
  \bibinfo{person}{Matthieu Rivain}.} \bibinfo{year}{2013}\natexlab{}.
\newblock \showarticletitle{{Masking against Side-Channel Attacks: A Formal
  Security Proof}}. In \bibinfo{booktitle}{{\em 2013 Annual International
  Conference on the Theory and Applications of Cryptographic Techniques
  (EUROCRYPT '13)}}. \bibinfo{address}{Berlin, Heidelberg},
  \bibinfo{pages}{142--159}.
\newblock


\bibitem[\protect\citeauthoryear{Rivain}{Rivain}{2009}]%
        {RivainSAC08}
\bibfield{author}{\bibinfo{person}{Matthieu Rivain}.}
  \bibinfo{year}{2009}\natexlab{}.
\newblock \bibinfo{booktitle}{{\em On the Exact Success Rate of Side Channel
  Analysis in the Gaussian Model}}.
\newblock \bibinfo{publisher}{Springer Berlin Heidelberg},
  \bibinfo{address}{Berlin, Heidelberg}, \bibinfo{pages}{165–183}.
\newblock


\bibitem[\protect\citeauthoryear{Rotem, Naveh, Ananthakrishnan, Weissmann, and
  Rajwan}{Rotem et~al\mbox{.}}{2012}]%
        {DBLP:journals/micro/RotemNAWR12}
\bibfield{author}{\bibinfo{person}{Efraim Rotem}, \bibinfo{person}{Alon Naveh},
  \bibinfo{person}{Avinash Ananthakrishnan}, \bibinfo{person}{Eliezer
  Weissmann}, {and} \bibinfo{person}{Doron Rajwan}.}
  \bibinfo{year}{2012}\natexlab{}.
\newblock \showarticletitle{Power-Management Architecture of the Intel
  Microarchitecture Code-Named Sandy Bridge}.
\newblock \bibinfo{journal}{{\em {IEEE} Micro\/}} \bibinfo{volume}{32},
  \bibinfo{number}{2} (\bibinfo{year}{2012}), \bibinfo{pages}{20--27}.
\newblock
\showDOI{%
\url{https://doi.org/10.1109/MM.2012.12}}


\bibitem[\protect\citeauthoryear{Saab, Rohatgi, and Hampel}{Saab
  et~al\mbox{.}}{2016}]%
        {cryptoeprint:2016/700}
\bibfield{author}{\bibinfo{person}{Sami Saab}, \bibinfo{person}{Pankaj
  Rohatgi}, {and} \bibinfo{person}{Craig Hampel}.}
  \bibinfo{year}{2016}\natexlab{}.
\newblock \bibinfo{title}{Side-Channel Protections for Cryptographic
  Instruction Set Extensions}.
\newblock \bibinfo{howpublished}{Cryptology ePrint Archive, Paper 2016/700}.
  (\bibinfo{year}{2016}).
\newblock
\showURL{%
\url{https://eprint.iacr.org/2016/700}}
\newblock
\shownote{\url{https://eprint.iacr.org/2016/700}.}


\bibitem[\protect\citeauthoryear{Schneider and Moradi}{Schneider and
  Moradi}{2015}]%
        {tobias-leakage-assessment}
\bibfield{author}{\bibinfo{person}{Tobias Schneider} {and}
  \bibinfo{person}{Amir Moradi}.} \bibinfo{year}{2015}\natexlab{}.
\newblock \showarticletitle{Leakage Assessment Methodology}. In
  \bibinfo{booktitle}{{\em Cryptographic Hardware and Embedded Systems -- CHES
  2015}}. \bibinfo{publisher}{Springer Berlin Heidelberg},
  \bibinfo{address}{Berlin, Heidelberg}, \bibinfo{pages}{495--513}.
\newblock


\bibitem[\protect\citeauthoryear{Schöne, Ilsche, Bielert, Velten, Schmidl, and
  Hackenberg}{Schöne et~al\mbox{.}}{2021}]%
        {energy-efficiency-amd-zen2}
\bibfield{author}{\bibinfo{person}{Robert Schöne}, \bibinfo{person}{Thomas
  Ilsche}, \bibinfo{person}{Mario Bielert}, \bibinfo{person}{Markus Velten},
  \bibinfo{person}{Markus Schmidl}, {and} \bibinfo{person}{Daniel Hackenberg}.}
  \bibinfo{year}{2021}\natexlab{}.
\newblock \showarticletitle{Energy Efficiency Aspects of the AMD Zen 2
  Architecture}. In \bibinfo{booktitle}{{\em 2021 IEEE International Conference
  on Cluster Computing (CLUSTER)}}. \bibinfo{pages}{562--571}.
\newblock
\showDOI{%
\url{https://doi.org/10.1109/Cluster48925.2021.00087}}


\bibitem[\protect\citeauthoryear{Standaert}{Standaert}{2017}]%
        {Standaert2017HowT}
\bibfield{author}{\bibinfo{person}{François-Xavier Standaert}.}
  \bibinfo{year}{2017}\natexlab{}.
\newblock \bibinfo{title}{How (not) to Use Welch's T-test in Side-Channel
  Security Evaluations}.
\newblock \bibinfo{howpublished}{Cryptology ePrint Archive, Paper 2017/138}.
  (\bibinfo{year}{2017}).
\newblock
\showURL{%
\url{https://eprint.iacr.org/2017/138}}


\bibitem[\protect\citeauthoryear{Wang, Paccagnella, He, Shacham, Fletcher, and
  Kohlbrenner}{Wang et~al\mbox{.}}{2022}]%
        {wan2022hertzbleed}
\bibfield{author}{\bibinfo{person}{Yingchen Wang}, \bibinfo{person}{Riccardo
  Paccagnella}, \bibinfo{person}{Elizabeth~Tang He}, \bibinfo{person}{Hovav
  Shacham}, \bibinfo{person}{Christopher~W. Fletcher}, {and}
  \bibinfo{person}{David Kohlbrenner}.} \bibinfo{year}{2022}\natexlab{}.
\newblock \showarticletitle{Hertzbleed: Turning Power {Side-Channel} Attacks
  Into Remote Timing Attacks on x86}. In \bibinfo{booktitle}{{\em 31st USENIX
  Security Symposium (USENIX Security 22)}}. \bibinfo{publisher}{USENIX
  Association}, \bibinfo{address}{Boston, MA}, \bibinfo{pages}{679--697}.
\newblock


\bibitem[\protect\citeauthoryear{Wikichip}{Wikichip}{2021}]%
        {amd-pbo}
\bibfield{author}{\bibinfo{person}{Wikichip}.} \bibinfo{year}{2021}\natexlab{}.
\newblock \bibinfo{title}{Precision Boost Overdrive (PBO) - AMD}.
\newblock \bibinfo{howpublished}{\url{https://en.wikichip.org/wiki/amd/pbo}}.
  (\bibinfo{year}{2021}).
\newblock
\newblock
\shownote{Accessed: 2022-07-25.}


\bibitem[\protect\citeauthoryear{Wu and Picek}{Wu and Picek}{2020}]%
        {wu2020remove}
\bibfield{author}{\bibinfo{person}{Lichao Wu} {and} \bibinfo{person}{Stjepan
  Picek}.} \bibinfo{year}{2020}\natexlab{}.
\newblock \showarticletitle{Remove some noise: On pre-processing of
  side-channel measurements with autoencoders}.
\newblock \bibinfo{journal}{{\em IACR Transactions on Cryptographic Hardware
  and Embedded Systems\/}} (\bibinfo{year}{2020}), \bibinfo{pages}{389--415}.
\newblock


\bibitem[\protect\citeauthoryear{Yan, Guo, Chen, and Mei}{Yan
  et~al\mbox{.}}{2015}]%
        {Yan2015ASO}
\bibfield{author}{\bibinfo{person}{Lin Yan}, \bibinfo{person}{Yao Guo},
  \bibinfo{person}{Xiangqun Chen}, {and} \bibinfo{person}{Hong Mei}.}
  \bibinfo{year}{2015}\natexlab{}.
\newblock \showarticletitle{A Study on Power Side Channels on Mobile Devices}.
  In \bibinfo{booktitle}{{\em Proceedings of the 7th Asia-Pacific Symposium on
  Internetware}}. \bibinfo{pages}{30–38}.
\newblock


\end{thebibliography}
\end{document}